\documentclass[letterpaper,twocolumn,prl,aps,showpacs,floatfix, 
superscriptaddress]{revtex4-2}
\usepackage{bbold}
\usepackage{amsmath,amsfonts,amsmath,mathtools,bbm,bm}
\usepackage{graphicx}
\usepackage{braket}
\usepackage[colorlinks,linkcolor=blue,urlcolor=blue,citecolor=blue]{hyperref}
\usepackage[all]{hypcap}
\usepackage[utf8]{inputenc}

\usepackage{blindtext}
\usepackage{enumitem}
\usepackage{xcolor}
\usepackage[normalem]{ulem}
\usepackage{float}
\setcitestyle{numbers,square}

\usepackage[normalem]{ulem}

\newcommand{\new}[1]{\textcolor{black}{#1}}

\def\({\left (}
\def\){\right )}

\graphicspath{{Figure/PNG/}{Figure/PDF/}{Figure/EPS/}{Figure/TEX/}{Figure/}}

\newcommand\mydots{\hbox to 1em{.\hss.\hss.}}

\begin{document}

\title{Eigenstate Thermalization Hypothesis correlations via non-linear Hydrodynamics}

\author{Jiaozi Wang}\email{jiaowang@uos.de}
\affiliation{University of Osnabr\"uck, Department of Mathematics/Computer Science/Physics, D-49076 Osnabr\"uck, Germany}

\author{Ruchira Mishra}
\affiliation{Kadanoff Center for Theoretical Physics \& James Franck Institute, University of Chicago, Chicago, IL 60637, USA}

\author{Tian-Hua Yang}
\affiliation{Department of Physics, Princeton University, Princeton, NJ 08544, USA}

\author{Luca V. Delacrétaz}
\affiliation{Kadanoff Center for Theoretical Physics \& James Franck Institute, University of Chicago, Chicago, IL 60637, USA}

\author{Silvia Pappalardi}
\email{pappalardi@thp.uni-koeln.de}
\affiliation{Institut f\"ur Theoretische Physik, Universit\"at zu K\"oln, Z\"ulpicher Straße 77, 50937 K\"oln, Germany}

\date{\today}

\begin{abstract} 
The thermalizing dynamics of many-body systems is often described through the lens of the Eigenstate Thermalization Hypothesis (ETH). ETH postulates that the statistical properties of observables, when expressed in the energy eigenbasis, are described by smooth functions, that also describe correlations among the matrix elements. However, the form of these functions is usually left undetermined, constituting a key missing component of the ETH framework.
In this work, we investigate the structure of such smooth functions by focusing on their Fourier transform, recently identified as free cumulants.
Using non-linear hydrodynamics, we provide a prediction for the universal scaling of the late-time behavior of time-ordered free cumulants in the thermodynamic limit.
The prediction is further corroborated by large-scale numerical simulations of several non-integrable one-dimensional spin models
which exhibit diffusive transport behavior. Good agreement is observed in both infinite and finite-temperature regimes and for a collection of local observables. Our results indicate that the smooth multi-point correlation functions within the ETH framework admit a universal hydrodynamic description at low frequencies.

\end{abstract}

\maketitle

{\bf{Introduction}} - 
Understanding universal phenomena appearing in the classical and quantum dynamics of generic many-body systems is a long-standing theoretical challenge. A distinctive ubiquitous feature of classical dynamics is \emph{hydrodynamics}, which describes the late-time behavior of thermalizing systems in terms of the fluctuations of conserved densities. Effective field theories for fluctuating hydrodynamics (EFT) \cite{Martin:1973zz,Spohn_2014,RevModPhys.87.593,Crossley:2015evo}
have proven powerful in characterizing the long-time evolution of these systems. In the simplest situations, hydrodynamics manifests as diffusion: at large scales, observables overlapping with the conserved densities display a slow power-law decay, reflecting this universal diffusive nature. Beyond classical systems, the principles of hydrodynamics extend into the quantum domain since they describe the long-time dynamics of quantum many-body systems \cite{KADANOFF1963419, PhysRevB.73.035113, PhysRevA.89.053608, Michailidis:2023mkd}, including integrable ones \cite{castroalvaredo2016emergent, bertini2016transport, doyon2020lecture, alba2021generalized, doyon2024generalized}, and have been observed across experimental platforms from complex materials to cold atoms \cite{Schemmer2019Generalized, Malvania2021Generalized, Bouchoule2022Generalized}.

In quantum systems, another universal aspect of thermalization is the \emph{Eigenstate Thermalization Hypothesis} (ETH) \cite{deutsch1991quantum,srednicki1994chaos}. 
If the system is non-integrable, ETH conjectures that the matrix elements $A_{\alpha_1 \alpha_2} = \bra{\alpha_1} \hat{A} \ket{\alpha_2}$ of a physical local observable $\hat{A}$ in the energy eigenbasis ($\hat H\ket{\alpha_i}=E_{\alpha_i} \ket{\alpha_i}$) shall behave as pseudorandom matrices, with smooth statistical properties \cite{srednicki1996thermal,srednicki1999approach}.  Intensive numerical studies \cite{rigol2008thermalization,eth-num-PhysRevB.99.155130,eth-num-PhysRevE.100.062134,eth-num-PhysRevE.82.031130,eth-num-PhysRevE.87.012118,eth-num-PhysRevE.89.042112,eth-num-PhysRevE.90.052105,eth-num-PhysRevE.91.012144,eth-num-PhysRevE.93.032104,eth-num-PhysRevE.96.012157,Delacretaz:2022ojg} have supported the validity of ETH in a wide range of local many-body systems, making ETH a well-established paradigm for thermalization. 

Recently, several studies have focused on \emph{correlations among these matrix elements} \cite{foini2019eigenstate, chan2019eigenstate, murthy2019bounds, richter2020eigenstate, brenes2021out, wang2021eigenstate, dymarsky2022bound}. This extended formulation of ETH, or ``full ETH'' \cite{foini2019eigenstate}, generalizes the standard approach to encompass multi-point correlations.  Specifically, it states that the statistical averages of products of matrix elements with repeated indices factorize, while with distinct indices,
$\alpha_1 \ne \alpha_2 \ne \dots \ne \alpha_n$ satisfy
\begin{equation}\label{eth1}
\overline{A_{\alpha_1 \alpha_2}A_{\alpha_2 \alpha_3}\dots A_{\alpha_n \alpha_1}} = D_{E^+}^{1-n} F_{e^+}^{(n)} \left(\omega_{\alpha_1\alpha_2},\dots, \omega_{\alpha_{n-1}\alpha_n}\right)
\end{equation}
where $D_{E^+}$ is the density of states at the average energy $E^+ \equiv (E_{\alpha_1}+\cdots + E_{\alpha_n})/n$. Here $F_{e^+}^{(n)}$ are \emph{smooth functions} of the average energy density $e^+ = E^+/L $ and the eigenenergy differences $\omega_{\alpha_1\alpha_2}=E_{\alpha_1}-E_{\alpha_2}$ \cite{footnote2}. These ``ETH functions'' (also known as on-shell correlations \cite{pappalardi2024microcanonical}) will be the focus of this paper.

Recent work has revealed connections between ETH and free probability theory \cite{pappalardi2022eigenstate} -- an extension of standard probability theory to non-commuting objects \cite{Voiculescu1985Symmetries, Mingo2017Free,Speicher1997Free} -- by establishing that the ETH functions are related to ``free cumulants''. Free cumulants are similar to connected correlation functions but differ from these by the combinatorics of non-crossing partitions, as reviewed below. 
This motivated growing interest in the emergence of free probability in the description of high-order correlations in chaotic quantum dynamics \cite{Pappalardi2025Full, wang2024emergence, Jindal2024Generalized, Fava2025Designs, Fritzsch2025Microcanonical, Chen2025FreeIndependence, Vallini2024LongTime, ampelogiannis2024cluster, Camargo2025QuantumSignatures, alves2025probes}. However, despite this progress, the physical content and universal structure of the ETH free cumulants correlations in chaotic local systems remain largely unexplored.
While ETH postulates the existence of the smooth functions $F_{e^+}^{(n)}(\vec{\omega})$ in Eq.~\eqref{eth1} for any order $n$, it does not predict their specific form, which depends in principle on the details of the underlying physics. What determines their shape in generic thermalizing systems? Do they obey any hierarchy?

In this work, we demonstrate that hydrodynamics controls the ETH functions \eqref{eth1} at small frequencies $\omega_{\alpha_i\alpha_{i+1}}$ for a large class of local operators $\hat A$. While this was known for two-time correlations \cite{Delacretaz:2020nit,capizzi2024hydrodynamics}, we extend this analysis to multi-time functions using the full non-linear structure of hydrodynamic effective field theories \cite{RevModPhys.87.593,Crossley:2015evo,delacretaz2024nonlinear}. For simplicity, we focus on systems with a single conserved quantity, such as energy, though our results can be easily generalized. 
We build on the hydrodynamic analysis of classical cumulants, which are suppressed with increasing powers of time at long times \cite{delacretaz2024nonlinear}, thus establishing a hierarchy.
This allows us to predict the late-time behavior of time-ordered free cumulants in the thermodynamic limit: \emph{even order time-ordered free cumulants always factorize in terms of two-point functions at sufficiently long times.} 
These results are general and apply to all observables that overlap with a conserved density or its derivatives. 
We test our predictions using large-scale numerics on three one-dimensional spin models, using a combination of exact diagonalization and dynamical quantum typicality (DQT) \cite{DQT-Gemmer}. The good agreement between hydrodynamic predictions and numerical results for finite-size quantum dynamics at long times
constitutes another important result of our work -- hydrodynamic tails are particularly challenging to observe in quantum simulations, as discussed in recent quench dynamics studies \cite{maceira2024thermalization, matthies2024thermalization}. Finally, we discuss the crossover from polynomial to late-time exponential decay due to the finite size of the lattice,  providing further validation of hydrodynamic behavior.

\medskip

{\bf{ETH smooth functions as free cumulants}} - 
The formalism of the full ETH in Eq.~\eqref{eth1} is greatly simplified using free probability theory and, in particular, the definition of 
 \emph{free cumulants} $\kappa_n(t_{n-1}, \dots, t_0)$ \cite{Speicher1997Free}. These are connected correlation functions of multi-time equilibrium expectation values, such as the canonical one, i.e. $\langle \cdot \rangle = \text{tr}(e^{-\beta H}\cdot) /Z$ with $Z=\text{tr}(e^{-\beta H})$. Free cumulants are defined implicitly using the combinatorics of non-crossing partitions as \cite{noncrossing}
\begin{equation}
    \label{eq_freecumbeta}
    \langle \hat A(t_{n-1}) \hat A(t_{n-2}) \dots \hat A(t_0) \rangle = \sum_{\pi\in NC(q)} \kappa_\pi \left ( t_{n-1},  \dots  t_0\right ) .
\end{equation}
Considering an observable with zero average $\langle \hat{A}\rangle=0$, the first few free cumulants are defined as \cite{first-moment},
\begin{subequations}
\label{free_cum_def}
\begin{gather}
\kappa_1(0)  \equiv \langle \hat{A}\rangle = 0 \ , 
\kappa_2(t_1, 0) \equiv \langle \hat{A}(t_1) \hat{A}\rangle \ ,\\
\kappa_3(t_2, t_1, 0)  \equiv \langle \hat{A}(t_2) \hat{A}(t_1) \hat{A}\rangle \ ,\\
\kappa_4(t_3, t_2, t_1, 0)  \equiv \langle \hat{A}(t_3) \hat{A}(t_2) \hat{A}(t_1) \hat{A}\rangle \\
 \quad - \braket{\hat{A}(t_3) \hat{A}(t_2) }\braket{\hat{A}(t_1) \hat{A} } 
- \braket{\hat{A}(t_3) \hat{A} }\braket{\hat{A}(t_2) \hat{A}(t_1) } \ . \nonumber
\end{gather}
\label{eq-fc}
\end{subequations}

When ETH applies, then one can show that 
 free cumulants for large $L$ are given by the so-called ETH free cumulants \cite{pappalardi2022eigenstate}
\begin{align}
\label{eq_free_cum_eth}
\kappa_n(t_{n-1}, \dots, t_0)
& =  \kappa^{\rm ETH}_n(t_{n-1}, \dots, t_0)
+ \mathcal O(L^{-1}) \ ,
\end{align}
which are defined as sums over different indices that directly encode smooth ETH correlations in Eq.~\eqref{eth1}, i.e.
\begin{align}
\label{eq_Kn_eth}
    \kappa^{\rm ETH}_n(\vec t)  
& = \frac{1}{Z} \sum_{\alpha_1\ne\dots \ne \alpha_n}  e^{-\beta E_{\alpha_1}}e^{i \vec \omega \cdot \vec t}  A_{\alpha_1 \alpha_2} \dots A_{\alpha_n \alpha_1}  \nonumber \\
& = \text{FT} \left [F_{e_\beta}^{(n)}(\vec \omega) e^{-\beta \vec \omega \cdot \vec \ell_n} \right ]    \ .
\end{align}
Here $\vec t = (t_{n-1}, \dots t_1)$, $\vec \omega=(\omega_{n-1}, \dots, \omega_1) $ with $\omega_n = \omega_{\alpha_{n-1}\alpha_{n-2}}$ and $\ell_n=(n-1, n-2, \dots, 1)/n$ makes explicit the fluctuation-dissipation-theorem at higher-order \cite{haehl2017thermal}, with ${e_\beta}=\langle \hat H\rangle/L$ the equilibrium energy. 

The validity of Eq.~\eqref{eq_free_cum_eth} is based on statistical arguments on the scaling of the matrix elements and the smoothness of the ETH free cumulants, and it has been verified numerically in few and many-body systems \cite{Pappalardi2025Full, Fritzsch2025Microcanonical, Vallini2024LongTime, alves2025probes}. Within ETH, correlations are smooth yet undetermined functions whose shape in frequency encodes all the dynamical behavior of a given observable. In what follows, we will see how hydrodynamics imposes constraints on their long-time (low-frequency) properties. 

Finally, let us recall that free cumulants in Eq.\eqref{eq_freecumbeta}, which play an important role in ETH, differ from more familiar connected correlation functions or ``classical cumulants'' $\langle \hat A(t_{n-1})  \dots \hat A(t_0)\rangle_c$. 
The latter also admits a combinatorial implicit definition similar to Eq.~\eqref{eq_freecumbeta}, but in terms of all set partitions \cite{footnote1}.
Hence, their difference stems from the presence of crossing partitions, which is known at all orders and first arise at the fourth one (See the End Matter for more details): 
\begin{align}\label{eq-cc-fc4}
\kappa_4(t_3, t_2, t_1, 0) & = \langle \hat A(t_3) \hat A(t_2) \hat A(t_1) \hat A\rangle_c  \nonumber \\
& \quad + \braket{\hat A(t_3) \hat A(t_1) }_c \braket{\hat A(t_2) \hat A }_c\ .
\end{align}

{\bf{Non-linear hydrodynamics}} - 
We consider a prototypical non-integrable system with local interactions, where the  Hamiltonian $\hat H = \sum_{i=1}^N \hat h_i$ is the only conserved quantity.
The late-time dynamics is expected to feature a diffusion of the energy density $\hat h_i$. The two-point function, capturing linear response, assumes a universal form at late times in diffusive systems:
\begin{equation}\label{eq_diff2pt}
\langle\hat{h}_{x}(t)\hat{h}_{0}(0)\rangle_{c}=\frac{C}{|t|^{d/2}}e^{\frac{-x^{2}}{4D|t|}}\left[1+\mathcal{O}(\tfrac{1}{t^{d/2}})+\mathcal{O}(\tfrac{1}{t})\right],
\end{equation}
where the constant $C=\frac{c_V/\beta^2}{(4\pi D)^{d/2}}$ depends on the diffusivity $D$ and on the specific heat $c_V\equiv -\beta^2 de/d\beta$. The $\mathcal O (\tfrac1{t^{d/2}})$ and $\mathcal O (\tfrac1{t})$ terms come from fluctuation corrections and higher derivative corrections respectively; the former dominate asymptotically in $d=1$ dimensions.

Notably, diffusive systems generically display non-linear response, the leading source of which is the density dependence of the diffusivity and susceptibility (or specific heat). 
Using the effective field theory of diffusion, Ref.~\cite{delacretaz2024nonlinear} presented a scaling prediction for the late-time behavior of the 
standard time-ordered connected $n$-point function, from now on called ``classical cumulants'' \cite{footnote1}.
In diffusive systems, the time-ordered classical cumulants of the energy densities of $n$ times with $t_{n-1}>\dots > t_{2}>t_1>0$ behave as
\begin{equation}
\label{hydro_nl0}
\langle \hat h_{x_{n-1}}(t_{n-1})\cdots \hat h_{x_{1}}(t_{1}) \hat h \rangle_c
= \frac C{\bar t^{(n-1)d/2}} g_n(\vec \tau, \vec y) + \cdots ,
\end{equation}
where the $\cdots$ denote relative corrections of the same order as in Eq.~\eqref{eq_diff2pt}. Here $g_n(\vec \tau, \vec y)$ involves several universal scaling functions which depend on the cross-ratio of times $\vec \tau = (t_{n-2}/ t_{n-1}, \dots t_2/t_1)$ and of the coordinates $\vec y = (y_i/\sqrt{Dt_i})$. The time dependence is expressed in terms of the geometric mean
${\bar t = [(t_{n-1}-t_{n-2})\dots (t_{2}-t_{1})(t_{1}-0)]^{\frac{1}{n-1}}}$. 

A similar logic governs the non-linear correlation functions of other local operators $\hat A$. The general strategy is to expand the operator into composite hydrodynamic fields, in an expansion in derivatives and fluctuations \cite{Delacretaz:2020nit,matthies2024thermalization}:
\begin{equation}\label{eq_general_op}
\hat A 
    = \alpha_1 \hat h(x_i) + \alpha_2 \partial_x \hat h(x_i) + \alpha_3 [\hat h(x_i)]^2 + \cdots\, , 
\end{equation}
where some coefficients $\alpha_n$ may vanish by symmetry. We emphasize that this equation does not hold as a microscopic operator equation, but rather as an operator equation at the level of the hydrodynamic EFT. The leading late time correlation functions of $A_i$ will then be determined by the most relevant term appearing on the right-hand side of Eq.~\eqref{eq_general_op}. Under diffusive scaling, $\partial_x\sim 1/t^{1/2}$. Eq.~\eqref{eq_diff2pt} further implies the scaling $h(x)\sim 1/t^{d/4}$. If the leading term in Eq.~\eqref{eq_general_op} then leads to a scaling $A\sim 1/t^{\delta_A}$, then non-linear correlators behave as
\begin{equation}
\label{hydro_nl}
\langle A_{x_{n-1}}(t_{n-1})
\cdots A_{x_{1}}(t_{1})A \rangle_c 
\sim \frac{1}{\bar t^{n\delta_A + (n-2)d/4}} \, .
\end{equation}
The additional suppression of $1/\bar t^{(n-2)d/4}$ arises from the fact that  non-Gaussianities are small (irrelevant) in the effective field theory of diffusion.
For the density $\hat A=\hat h_i$, which has scaling dimension $\delta_h = d/4$, one recovers \eqref{eq_general_op}. As another example, consider the energy current: $\hat A = \hat j_i$. Since it is odd under parity, $\alpha_{1,3}$ must vanish in \eqref{eq_general_op} and the most relevant hydrodynamic term is $\partial_x h$, which has dimension $\delta_{ j} = (d+2)/4$.

{\bf{ETH correlations via hydrodynamics}} - 
The hydrodynamic scaling in Eq.~\eqref{hydro_nl} shows that time-ordered multi-time correlation functions are organized according to their classical cumulants: the higher the order $n$, the faster the decay in time of the classical connected function. Importantly, this behavior also determines the behavior of the ETH free cumulants in Eq.~\eqref{eq_free_cum_eth}. 

We start by considering as an observable the energy density $\hat A = \hat h_i$, which, being local, is expected to obey ETH in non-integrable systems. Using the relation between free and classical cumulants in Eqs.~\eqref{eq-cc-fc} and the suppression of latter cumulants at long times [cf. Eq.~\eqref{hydro_nl}], we deduce that time-ordered free cumulants behave as
\begin{subequations}
\label{eq_eth_hydro}
\begin{gather}
\label{eq_eth_hydro_a}
\kappa_3(t_2, t_1, 0)  \sim \frac 1{\bar t^d}\\
\kappa_4(t_3, t_2, t_1, 0)  = \braket{h_{x_3}(t_3) h_{x_1}(t_1) }_c \braket{h_{x_2}(t_2) h_0(0) }_c + \mathcal O( \frac 1{\bar t^{3d/2}})
\label{eq_eth_hydro_b}
\end{gather}
\end{subequations}
where the correction in the last equation is given by Eq.~\eqref{hydro_nl}, with $\delta_h = d/4$ and $n=4$. This behavior can be translated as a prediction in the frequency domain
for three-point functions, as shown in the End Matter.
Eqs.~\eqref{eq_eth_hydro} place stringent constraints on the correlations among the matrix elements: These are not governed by independent new functions for every order $n$, but rather governed by universal hydrodynamic tails at long times. Furthermore, from Eq.~\eqref{eq:kappa-to-c}, even correlators asymptotically factorize in products of two-point functions (see, e.g., Eq.~\eqref{eq_eth_hydro_b}), and odd correlators factorize to products of two-point functions and one three-point function,  as
\begin{align}\label{eq_factorize}
    \kappa_{2m}(\vec t) & \simeq \frac{1}{\bar t^{dm/2}}\ ,\quad
     \kappa_{2m+1}(\vec t)  \simeq \frac{1}{\bar t^{(m+1)d/2}}
    \ .
\end{align}    
As for non-linear hydrodynamics in Eqs.~\eqref{eq_general_op}-\eqref{hydro_nl}, these results are supposed to apply to all observables that overlap with the energy density or its derivatives.

{\bf{Numerical evaluation}} - 
We expect the scaling behavior in Eq.~\eqref{eq_eth_hydro} to apply to any diffusive many-body system. 
We test this prediction on the one-dimensional mixed-field Ising model for spin $S=1$, governed by the Hamiltonian $\hat{H}=\sum_{i}\hat{h}_{i}$, where $\hat{h}_{i}=J\hat{s}_{i}^{z}\hat{s}_{i+1}^{z}+\frac{1}{2}g_{z}(\hat{s}_{i}^{z}+\hat{s}_{i+1}^{z})+\frac{1}{2}g_{x}(\hat{s}_{i}^{x}+\hat{s}_{i+1}^{x})$
denotes the energy density.
Here $\hat s_i^\mu$ are spin $1$ operators in the $\mu=x, z$ direction and we use periodic boundary conditions $\hat{s}_{L+1}^{\mu}=\hat{s}_{1}^{\mu}$. The parameters are fixed as $g_x=1.1$, $g_z=0.9$ and $J = 0.707$.
For spin $S=1/2$, the hydrodynamics of this model has been explored, e.g. in Refs.~\cite{leviatan2017quantum, klein2022time, rakowsky2022dissipation, Mori:2023qbd, artiaco2024efficient, maceira2024thermalization,jonas-Ising-10.21468/SciPostPhys.9.3.031,white-Ising-PhysRevB.110.134308}. We instead focus in the main text on the spin $S = 1$ case~\cite{numerics}, for which the hydrodynamic long-time tails become visible already at relatively small system sizes~\cite{capizzi2024hydrodynamics}.
We investigate numerically multi-time correlation functions up to $n=2,3,4$ for different local observables $\hat A$. 
The results are obtained with exact diagonalization (ED) and Dynamical Quantum Typicality (DQT) \cite{DQT-Gemmer}.
\begin{figure}[t]
	\centering
	\includegraphics[width=1.0 \linewidth]{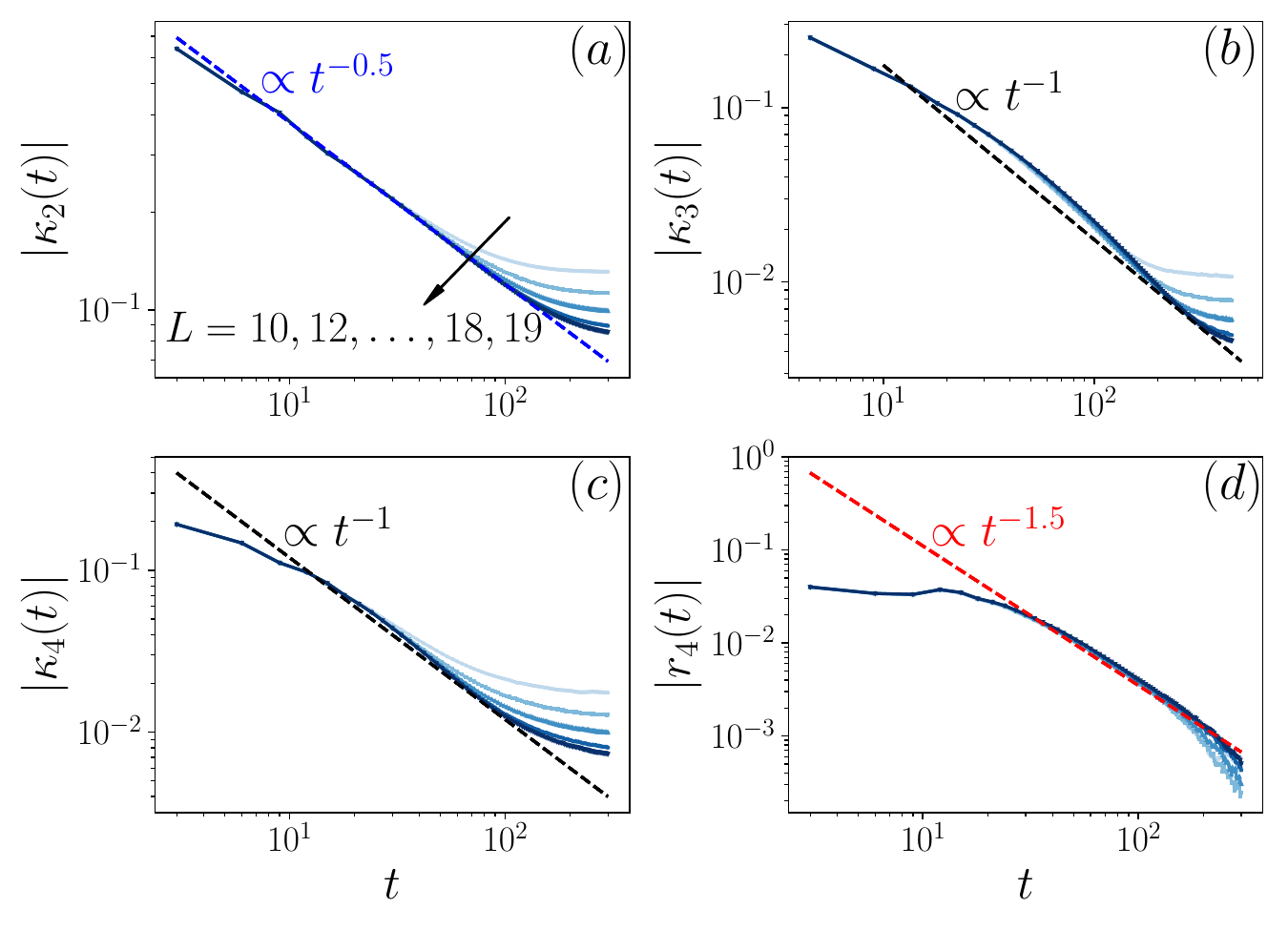}
	\caption{Free cumulants $\kappa_n(t)$ and classical cumulants $r_n(t)$ of the energy density $\hat A = \hat h_i$ versus time $t$ in the spin-1 Ising model: (a) $\kappa_2(t)$; (b) $\kappa_4(t)$; (c) $\kappa_3 (t)$ and (d) $r_4(t) = \kappa_4(t) - (\kappa_2(2t))^2$ for $L = 10,12,14,16,18,19$  (from light to dark blue), at infinite temperature $\beta = 0$. Blue, black and red dashed lines indicate the hydrodynamic predictions from Eq.~\eqref{eq_cumu_prediction}. Results are obtained from ED for $L=10$ and DQT for $L>10$ \cite{single-precesion}.}
	\label{fig:1}
\end{figure}

In the case of the energy density $\hat h_{x_i}=\hat h_0$ on the same site, Eqs.~\eqref{eq_eth_hydro} predict that the free cumulants shall scale at long-times as
\begin{subequations}
\label{eq_cumu_prediction}
\begin{align}
    \kappa_2(t) \equiv \kappa_2(t, 0) & \simeq  c_2 t^{-1/2} \ , 
    \label{eq_cumu_prediction_a}\\
    \kappa_3(t) \equiv \kappa_3(2t, t, 0) & \simeq c_3 t^{-1} \ , 
    \label{eq_cumu_prediction_b}\\
    \kappa_4(t) \equiv \kappa_4(3t, 2t, t, 0) & \simeq  (c_2 2t^{-1/2})^2 + c_4t^{-3/2} \ .
    \label{eq_cumu_prediction_c}
\end{align}    
\end{subequations}
To illustrate the predicted long-time scaling, we plot in Fig.~\ref{fig:1}(a-c) the free cumulants defined in Eqs.~\eqref{eq_cumu_prediction} as a function of time at infinite temperature $\beta=0$. As the plot clearly shows, by increasing system size up to $L=19$, the data converge toward the hydrodynamic prediction, represented by a dashed line in the plots.  To study the subleading correction, in panel (d) we subtract the factorized result from the fourth free cumulant, i.e. 
\begin{equation}
    \label{classical_cumu}
    r_4(t) \equiv\langle \hat A(3t) \hat A(2t) \hat A(t)\hat A\rangle_c= \kappa_4(t) - (\kappa_2(2t))^2 \ ,
\end{equation}
which corresponds to the fourth-order classical cumulant. Indeed, the data shows an excellent agreement with the prediction $r_4(t) \sim t^{-3/2}$, see also Eq.~\eqref{hydro_nl}. 
Higher-order non-linearities were previously studied numerically for a classical system \cite{delacretaz2024nonlinear} and a quantum Floquet system \cite{Michailidis:2023mkd}, but had to the best of our knowledge not been observed and compared to hydrodynamic predictions in a quantum Hamiltonian system, where ETH is expected to apply. In the limit of equal times, non-linear response reduces to ``full counting statistics'' which has received more attention; see, e.g., \cite{PhysRevB.51.4079} for studies in single-particle diffusive systems and  \cite{PhysRevLett.94.030601} in many-body systems.

\begin{figure}[t]
	\centering
	\includegraphics[width=1.0 \linewidth]{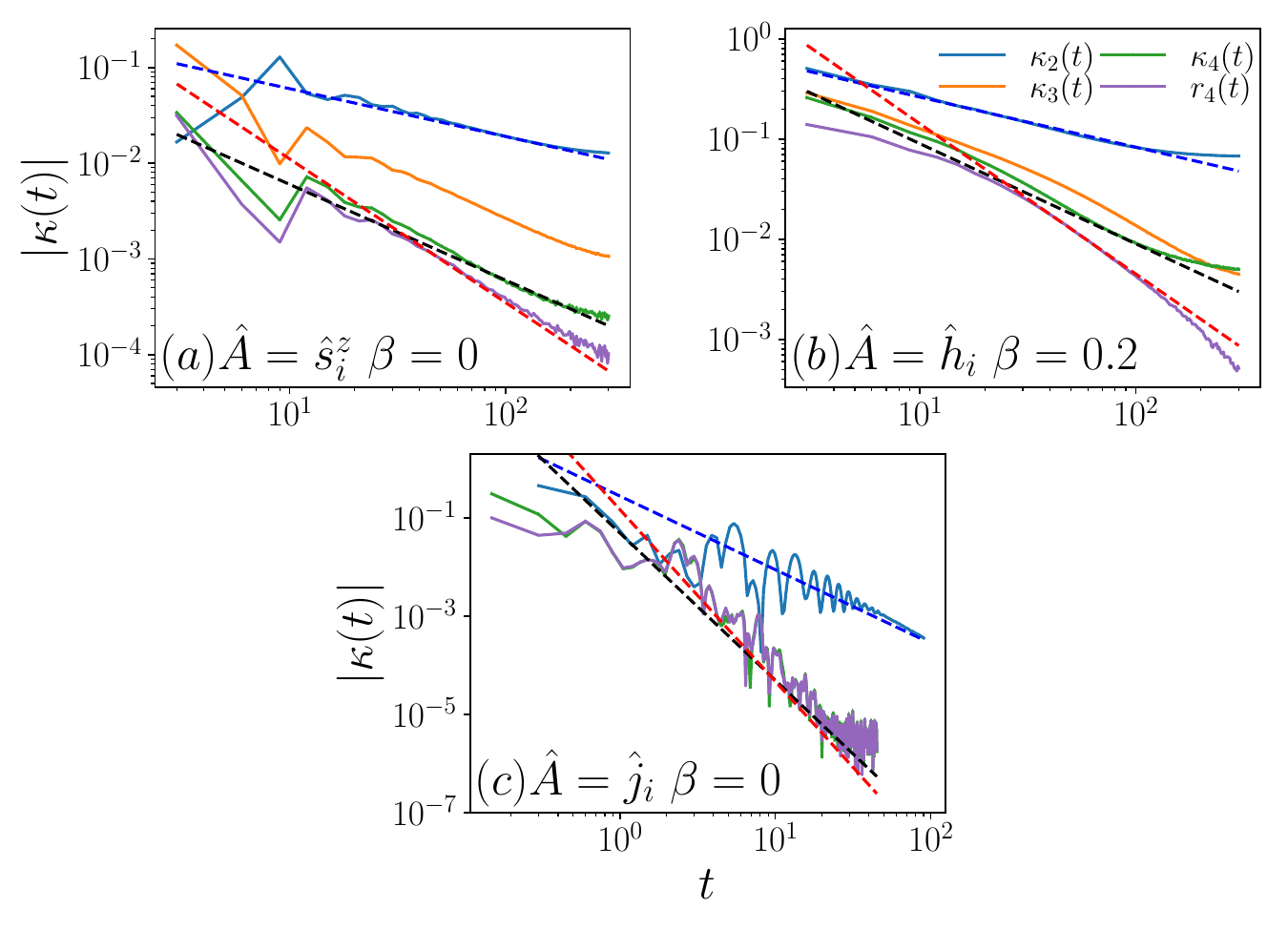}
	 \caption{  Free cumulants $\kappa_n (t)$ and classical cumulants $r_n(t)$
  for different operators and temperatures: (a) ${\hat A} = \hat{s}^z_i,\ \beta = 0$;
(b) ${\hat A} = \hat{h}_i,\ \beta = 0.2$, and (c) ${\hat A} = \hat{j}_i,\ \beta = 0$ for system  size $L=19$. Blue, black and red dashed lines indicate the hydrodynamic predictions $\propto t^{-1/2}, t^{-1},t^{-3/2}$ in (a, b) and $\propto t^{-3/2}, t^{-3},t^{-7/2}$ in (c). }
	\label{fig:2}
\end{figure}

To show the generality of our predictions, we test different temperature ranges and local observables $\hat A$ in Fig.\ref{fig:2}, where we plot the second, third and fourth free cumulants and the classical one and their hydrodynamical predictions represented by the dashed lines. 
In Fig.\ref{fig:2}(a-b), we consider the energy density $\hat A = \hat h_i$ at finite $\beta=0.2$ and a single site spin magnetization along $z$, $\hat A = \hat s_i^z$ at $\beta=0$ at $L=19$. Since the latter observable overlaps with the energy density, both the choices follow the hydrodynamic decay in Eqs.~\eqref{eq_cumu_prediction}, as shown by the good agreement in the plot. In Fig.\ref{fig:2}(c), we show the local energy current $\hat A= \hat{j}_{i}=Jg_{x}\hat{s}_{i}^{y}(\hat{s}_{i+1}^{z}-\hat{s}_{i-1}^{z})$, which overlaps instead with $\nabla h$, thus hydrodynamics predicts $\kappa_2(t) \sim t^{-3/2}$, $\kappa_4(t) \sim t^{-3}$ and $r_4(t) \sim t^{-7/2}$ ($\kappa_3(t) = 0$ due to reflection symmetry).  
The numerical results agree well with these predictions, although it is difficult to distinguish the last two power-law decays, $t^{-3}$ and $t^{-7/2}$\cite{further-numerics}.

{\bf Late time exponential decay--} Power-law decay in systems with conservation laws follows from an accumulation of slower and slower hydrodynamic `quasinormal modes', with longer and longer wavelengths. In a finite size system, this process eventually terminates at the lifetime of the longest hydrodynamic mode -- this determines the global relaxation time (or Thouless time), given by $\tau_{\rm global} = (L/2\pi)^2/D$ in a diffusive system. After this time, locality no longer plays an important role and the system behaves like a single quantum mechanical object. Correlation functions are then expected to decay exponentially $\sim e^{-\Gamma t}$, with a rate set by this slowest lifetime $\Gamma = 1/\tau_{\rm global}$. The only remaining signature from hydrodynamics in this very late time regime is, therefore, through the dependence of $\Gamma$ on system size $L$ and temperature or couplings through the diffusivity $D$.

\begin{figure}[]
	\centering
	\includegraphics[width=1. \linewidth]{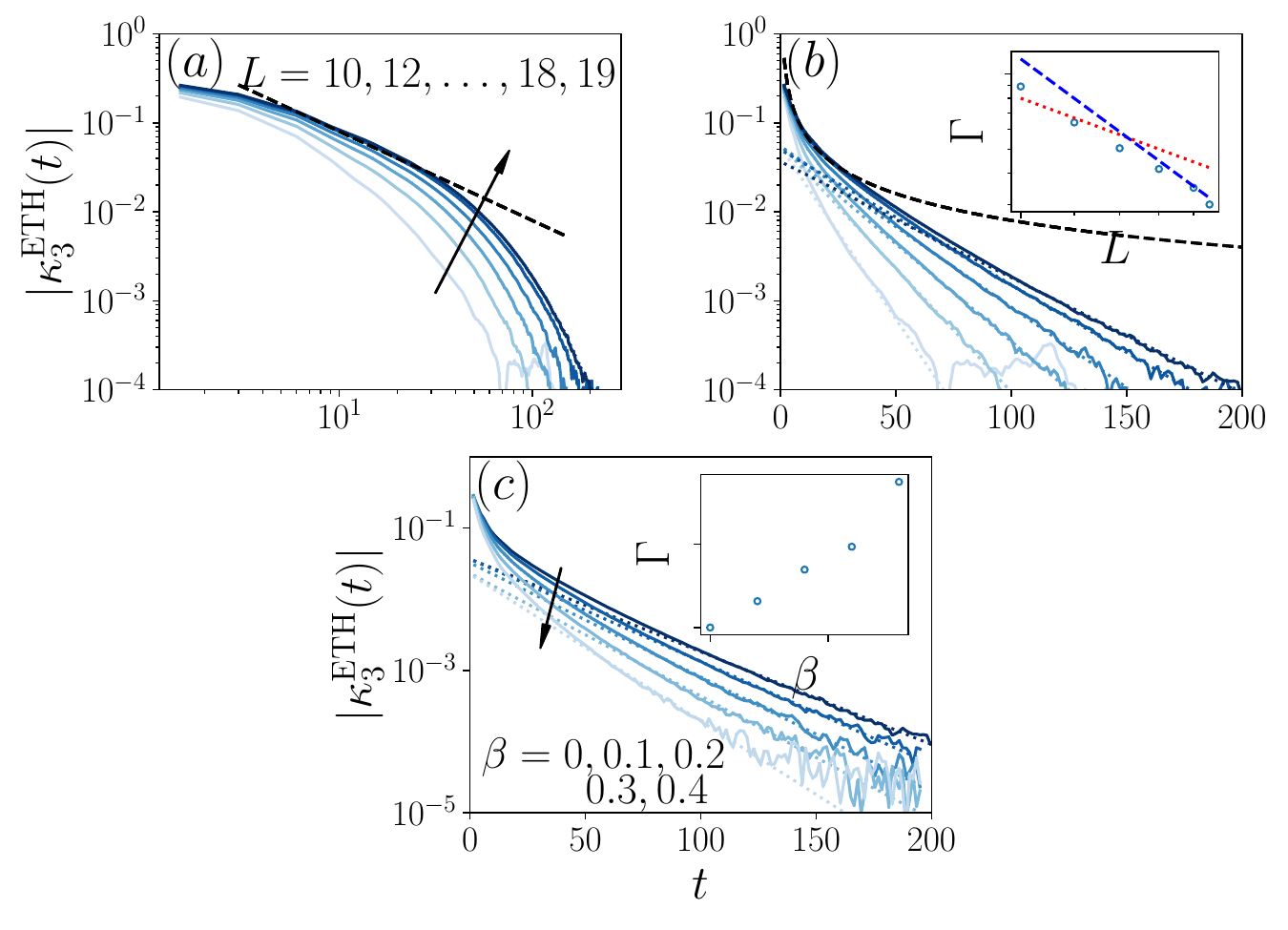}	\caption{ETH free cumulants $\kappa^\text{ETH}_3(t)$ as a function of time $t$ for the operator ${\hat A} = \hat{h}_i$ for (a, b) $\beta = 0$ with system sizes $L = 10, 12, 14, 16, 18, 19$ (from light to dark blue), and (c) $L = 19$ for $\beta = 0, 0.1, 0.2, 0.3, 0.4$ (from dark to light blue). The dashed line represents the power-law scaling $\propto t^{-1}$, and the dotted line corresponds to the exponential fit $\propto \exp(-\Gamma t)$. The exponential fitting is performed in the region where $\kappa_{3}^{\text{ETH}}(t)\in[10^{-2.5},10^{-4}]$, and the corresponding fitting parameters $\Gamma$ are shown in the insets of panel (b) on a double-logarithmic scale and panel (c) on a linear scale. }
	\label{fig:3}
\end{figure}

To cleanly extract the late time exponential decay, we remove the asymptotic value of correlators, visible in Fig.~\ref{fig:1}, by considering the ETH free cumulants in Eq.~\eqref{eq_Kn_eth}, given only by sums over different indices. Fig.~\ref{fig:3} shows the ETH free cumulant $\kappa_3^{\text{ETH}}(t)$ of the energy density $\hat{h}_i$ for different $L$ and temperatures $\beta$ ($\kappa_2^{\text{ETH}}(t)$ is shown in the End Matter). The intermediate time power-law behavior is seen to transition to exponential behavior $\sim e^{-\Gamma t}$ at later times. Fig.~\ref{fig:3}(b) shows that the decay rate $\Gamma$ decreases with increasing system size, and a $\Gamma \sim 1/L^2$ scaling can be roughly observed for $L\ge 16$, consistent with the hydrodynamic prediction \cite{gamma-scaling}.
Second, Fig.~\ref{fig:3}(c) shows that the decay rate $\Gamma$ decreases with increasing temperature, as observed in \cite{maceira2024thermalization}. This behavior follows, in fact, from hydrodynamic expectations: while higher temperatures lead to faster {\em local} thermalization, {\em global} thermalization—achieved through transport of conserved quantities across the system—becomes slower due to the shorter mean free path.

{\bf{Conclusion and Outlook}} - 
While free cumulants naturally arise in ETH, it is the classical cumulants that have a hierarchical structure in a hydrodynamic regime. This implies that free cumulants factorize, into products of two-point functions (or a product of two-point functions and one three point function for odd moments, see Eq.~\eqref{eq_factorize}). When ETH applies in its general form \eqref{eth1}, this provides a simple prediction for the statistics of matrix elements of local operators. These predictions are supported by numerics in three paradigmatic chaotic one-dimensional systems: correlators after the local equilibration time exhibit various power-law decays, with exponent depending on the operator and degree of the correlator. This behavior transitions to an exponential decay at late times, consistent with hydrodynamic predictions.

One aspect of the ETH functions that is not captured by current theories of hydrodynamics is out-of-time ordered correlators of $n\geq 4$ operators, whose Fourier transforms also contribute to Eq.~\eqref{eth1}. Capturing these would require extending hydrodynamic EFTs beyond their current scope. Another interesting direction would be to study higher cumulants in systems exhibiting anomalous transport such as super- or subdiffusion, and in integrable systems \cite{doi:10.1073/pnas.2106945118-nl-int} with the help of generalized hydrodynamics.

{\bf{Acknowledgment}} - 
We thank A. Rosch and A. L{\"a}uchli for helpful discussions. We also wish to thank A. Silva for initial discussions on the topic. 
SP acknowledges funding by the Deutsche
Forschungsgemeinschaft (DFG, German Research Foundation) under Projektnummer 277101999 - TRR 183
(project B02 and A03), and under Germany’s Excellence
Strategy - Cluster of Excellence Matter and Light for
Quantum Computing (ML4Q) EXC 2004/1 - 390534769.
JW acknowledges support from DFG, under Grant No. 531128043, as well as under Grant
No.\ 397107022, No.\ 397067869, and No.\ 397082825 within the DFG Research
Unit FOR 2692, under Grant No.\ 355031190. RM and LD are supported by an NSF award No.~PHY2412710.
Additionally, we greatly acknowledge computing time on the HPC3 at the University of Osnabr\"{u}ck, granted by the DFG, under Grant No. 456666331.

{\bf{Data availability}} - 
The data that support the findings of
this article are openly available \cite{data_wang}.
\bibliography{ref.bib}

\begin{thebibliography}{86}%
\makeatletter
\providecommand \@ifxundefined [1]{%
 \@ifx{#1\undefined}
}%
\providecommand \@ifnum [1]{%
 \ifnum #1\expandafter \@firstoftwo
 \else \expandafter \@secondoftwo
 \fi
}%
\providecommand \@ifx [1]{%
 \ifx #1\expandafter \@firstoftwo
 \else \expandafter \@secondoftwo
 \fi
}%
\providecommand \natexlab [1]{#1}%
\providecommand \enquote  [1]{``#1''}%
\providecommand \bibnamefont  [1]{#1}%
\providecommand \bibfnamefont [1]{#1}%
\providecommand \citenamefont [1]{#1}%
\providecommand \href@noop [0]{\@secondoftwo}%
\providecommand \href [0]{\begingroup \@sanitize@url \@href}%
\providecommand \@href[1]{\@@startlink{#1}\@@href}%
\providecommand \@@href[1]{\endgroup#1\@@endlink}%
\providecommand \@sanitize@url [0]{\catcode `\\12\catcode `\$12\catcode
  `\&12\catcode `\#12\catcode `\^12\catcode `\_12\catcode `\%12\relax}%
\providecommand \@@startlink[1]{}%
\providecommand \@@endlink[0]{}%
\providecommand \url  [0]{\begingroup\@sanitize@url \@url }%
\providecommand \@url [1]{\endgroup\@href {#1}{\urlprefix }}%
\providecommand \urlprefix  [0]{URL }%
\providecommand \Eprint [0]{\href }%
\providecommand \doibase [0]{https://doi.org/}%
\providecommand \selectlanguage [0]{\@gobble}%
\providecommand \bibinfo  [0]{\@secondoftwo}%
\providecommand \bibfield  [0]{\@secondoftwo}%
\providecommand \translation [1]{[#1]}%
\providecommand \BibitemOpen [0]{}%
\providecommand \bibitemStop [0]{}%
\providecommand \bibitemNoStop [0]{.\EOS\space}%
\providecommand \EOS [0]{\spacefactor3000\relax}%
\providecommand \BibitemShut  [1]{\csname bibitem#1\endcsname}%
\let\auto@bib@innerbib\@empty
\bibitem [{\citenamefont {Martin}\ \emph {et~al.}(1973)\citenamefont {Martin},
  \citenamefont {Siggia},\ and\ \citenamefont {Rose}}]{Martin:1973zz}%
  \BibitemOpen
  \bibfield  {author} {\bibinfo {author} {\bibfnamefont {P.~C.}\ \bibnamefont
  {Martin}}, \bibinfo {author} {\bibfnamefont {E.~D.}\ \bibnamefont {Siggia}},\
  and\ \bibinfo {author} {\bibfnamefont {H.~A.}\ \bibnamefont {Rose}},\
  }\bibfield  {title} {\bibinfo {title} {{Statistical Dynamics of Classical
  Systems}},\ }\href {https://doi.org/10.1103/PhysRevA.8.423} {\bibfield
  {journal} {\bibinfo  {journal} {Phys. Rev. A}\ }\textbf {\bibinfo {volume}
  {8}},\ \bibinfo {pages} {423} (\bibinfo {year} {1973})}\BibitemShut {NoStop}%
\bibitem [{\citenamefont {Spohn}(2014)}]{Spohn_2014}%
  \BibitemOpen
  \bibfield  {author} {\bibinfo {author} {\bibfnamefont {H.}~\bibnamefont
  {Spohn}},\ }\bibfield  {title} {\bibinfo {title} {Nonlinear fluctuating
  hydrodynamics for anharmonic chains},\ }\href
  {https://doi.org/10.1007/s10955-014-0933-y} {\bibfield  {journal} {\bibinfo
  {journal} {J. Stat. Phys.}\ }\textbf {\bibinfo {volume} {154}},\ \bibinfo
  {pages} {1191} (\bibinfo {year} {2014})}\BibitemShut {NoStop}%
\bibitem [{\citenamefont {Bertini}\ \emph {et~al.}(2015)\citenamefont
  {Bertini}, \citenamefont {De~Sole}, \citenamefont {Gabrielli}, \citenamefont
  {Jona-Lasinio},\ and\ \citenamefont {Landim}}]{RevModPhys.87.593}%
  \BibitemOpen
  \bibfield  {author} {\bibinfo {author} {\bibfnamefont {L.}~\bibnamefont
  {Bertini}}, \bibinfo {author} {\bibfnamefont {A.}~\bibnamefont {De~Sole}},
  \bibinfo {author} {\bibfnamefont {D.}~\bibnamefont {Gabrielli}}, \bibinfo
  {author} {\bibfnamefont {G.}~\bibnamefont {Jona-Lasinio}},\ and\ \bibinfo
  {author} {\bibfnamefont {C.}~\bibnamefont {Landim}},\ }\bibfield  {title}
  {\bibinfo {title} {Macroscopic fluctuation theory},\ }\href
  {https://doi.org/10.1103/RevModPhys.87.593} {\bibfield  {journal} {\bibinfo
  {journal} {Rev. Mod. Phys.}\ }\textbf {\bibinfo {volume} {87}},\ \bibinfo
  {pages} {593} (\bibinfo {year} {2015})}\BibitemShut {NoStop}%
\bibitem [{\citenamefont {Crossley}\ \emph {et~al.}()\citenamefont {Crossley},
  \citenamefont {Glorioso},\ and\ \citenamefont {Liu}}]{Crossley:2015evo}%
  \BibitemOpen
  \bibfield  {author} {\bibinfo {author} {\bibfnamefont {M.}~\bibnamefont
  {Crossley}}, \bibinfo {author} {\bibfnamefont {P.}~\bibnamefont {Glorioso}},\
  and\ \bibinfo {author} {\bibfnamefont {H.}~\bibnamefont {Liu}},\ }\bibfield
  {title} {\bibinfo {title} {Effective field theory of dissipative fluids},\
  }\href {https://doi.org/10.1007/JHEP09(2017)095} {\bibfield  {journal}
  {\bibinfo  {journal} {J. High Energy Phys.}} {\bibinfo {volume}
  {09}}\bibinfo  {number} { (2017)}\ \bibinfo {pages} {095}}\BibitemShut
  {NoStop}%
\bibitem [{\citenamefont {Kadanoff}\ and\ \citenamefont
  {Martin}(1963)}]{KADANOFF1963419}%
  \BibitemOpen
\bibfield  {number} {  }\bibfield  {author} {\bibinfo {author} {\bibfnamefont
  {L.~P.}\ \bibnamefont {Kadanoff}}\ and\ \bibinfo {author} {\bibfnamefont
  {P.~C.}\ \bibnamefont {Martin}},\ }\bibfield  {title} {\bibinfo {title}
  {Hydrodynamic equations and correlation functions},\ }\href
  {https://doi.org/https://doi.org/10.1016/0003-4916(63)90078-2} {\bibfield
  {journal} {\bibinfo  {journal} {Ann. Phys.}\ }\textbf {\bibinfo {volume}
  {24}},\ \bibinfo {pages} {419} (\bibinfo {year} {1963})}\BibitemShut
  {NoStop}%
\bibitem [{\citenamefont {Mukerjee}\ \emph {et~al.}(2006)\citenamefont
  {Mukerjee}, \citenamefont {Oganesyan},\ and\ \citenamefont
  {Huse}}]{PhysRevB.73.035113}%
  \BibitemOpen
  \bibfield  {author} {\bibinfo {author} {\bibfnamefont {S.}~\bibnamefont
  {Mukerjee}}, \bibinfo {author} {\bibfnamefont {V.}~\bibnamefont
  {Oganesyan}},\ and\ \bibinfo {author} {\bibfnamefont {D.}~\bibnamefont
  {Huse}},\ }\bibfield  {title} {\bibinfo {title} {Statistical theory of
  transport by strongly interacting lattice fermions},\ }\href
  {https://doi.org/10.1103/PhysRevB.73.035113} {\bibfield  {journal} {\bibinfo
  {journal} {Phys. Rev. B}\ }\textbf {\bibinfo {volume} {73}},\ \bibinfo
  {pages} {035113} (\bibinfo {year} {2006})}\BibitemShut {NoStop}%
\bibitem [{\citenamefont {Lux}\ \emph {et~al.}(2014)\citenamefont {Lux},
  \citenamefont {M\"uller}, \citenamefont {Mitra},\ and\ \citenamefont
  {Rosch}}]{PhysRevA.89.053608}%
  \BibitemOpen
  \bibfield  {author} {\bibinfo {author} {\bibfnamefont {J.}~\bibnamefont
  {Lux}}, \bibinfo {author} {\bibfnamefont {J.}~\bibnamefont {M\"uller}},
  \bibinfo {author} {\bibfnamefont {A.}~\bibnamefont {Mitra}},\ and\ \bibinfo
  {author} {\bibfnamefont {A.}~\bibnamefont {Rosch}},\ }\bibfield  {title}
  {\bibinfo {title} {Hydrodynamic long-time tails after a quantum quench},\
  }\href {https://doi.org/10.1103/PhysRevA.89.053608} {\bibfield  {journal}
  {\bibinfo  {journal} {Phys. Rev. A}\ }\textbf {\bibinfo {volume} {89}},\
  \bibinfo {pages} {053608} (\bibinfo {year} {2014})}\BibitemShut {NoStop}%
\bibitem [{\citenamefont {Michailidis}\ \emph {et~al.}(2024)\citenamefont
  {Michailidis}, \citenamefont {Abanin},\ and\ \citenamefont
  {Delacr\'etaz}}]{Michailidis:2023mkd}%
  \BibitemOpen
  \bibfield  {author} {\bibinfo {author} {\bibfnamefont {A.~A.}\ \bibnamefont
  {Michailidis}}, \bibinfo {author} {\bibfnamefont {D.~A.}\ \bibnamefont
  {Abanin}},\ and\ \bibinfo {author} {\bibfnamefont {L.~V.}\ \bibnamefont
  {Delacr\'etaz}},\ }\bibfield  {title} {\bibinfo {title} {Corrections to
  diffusion in interacting quantum systems},\ }\href
  {https://doi.org/10.1103/PhysRevX.14.031020} {\bibfield  {journal} {\bibinfo
  {journal} {Phys. Rev. X}\ }\textbf {\bibinfo {volume} {14}},\ \bibinfo
  {pages} {031020} (\bibinfo {year} {2024})}\BibitemShut {NoStop}%
\bibitem [{\citenamefont {Castro-Alvaredo}\ \emph {et~al.}(2016)\citenamefont
  {Castro-Alvaredo}, \citenamefont {Doyon},\ and\ \citenamefont
  {Yoshimura}}]{castroalvaredo2016emergent}%
  \BibitemOpen
  \bibfield  {author} {\bibinfo {author} {\bibfnamefont {O.~A.}\ \bibnamefont
  {Castro-Alvaredo}}, \bibinfo {author} {\bibfnamefont {B.}~\bibnamefont
  {Doyon}},\ and\ \bibinfo {author} {\bibfnamefont {T.}~\bibnamefont
  {Yoshimura}},\ }\bibfield  {title} {\bibinfo {title} {Emergent hydrodynamics
  in integrable quantum systems out of equilibrium},\ }\href
  {https://doi.org/10.1103/PhysRevX.6.041065} {\bibfield  {journal} {\bibinfo
  {journal} {Phys. Rev. X}\ }\textbf {\bibinfo {volume} {6}},\ \bibinfo {pages}
  {041065} (\bibinfo {year} {2016})}\BibitemShut {NoStop}%
\bibitem [{\citenamefont {Bertini}\ \emph {et~al.}(2016)\citenamefont
  {Bertini}, \citenamefont {Collura}, \citenamefont {Nardis},\ and\
  \citenamefont {Fagotti}}]{bertini2016transport}%
  \BibitemOpen
  \bibfield  {author} {\bibinfo {author} {\bibfnamefont {B.}~\bibnamefont
  {Bertini}}, \bibinfo {author} {\bibfnamefont {M.}~\bibnamefont {Collura}},
  \bibinfo {author} {\bibfnamefont {J.~D.}\ \bibnamefont {Nardis}},\ and\
  \bibinfo {author} {\bibfnamefont {M.}~\bibnamefont {Fagotti}},\ }\bibfield
  {title} {\bibinfo {title} {Transport in out-of-equilibrium xxz chains: Exact
  profiles of charges and currents},\ }\href
  {https://doi.org/10.1103/PhysRevLett.117.207201} {\bibfield  {journal}
  {\bibinfo  {journal} {Phys. Rev. Lett.}\ }\textbf {\bibinfo {volume} {117}},\
  \bibinfo {pages} {207201} (\bibinfo {year} {2016})}\BibitemShut {NoStop}%
\bibitem [{\citenamefont {Doyon}(2020)}]{doyon2020lecture}%
  \BibitemOpen
  \bibfield  {author} {\bibinfo {author} {\bibfnamefont {B.}~\bibnamefont
  {Doyon}},\ }\bibfield  {title} {\bibinfo {title} {Lecture notes on
  generalised hydrodynamics},\ }\href
  {https://doi.org/10.21468/SciPostPhysLectNotes.18} {\bibfield  {journal}
  {\bibinfo  {journal} {SciPost Phys. Lect. Notes}\ }\textbf {\bibinfo {volume}
  {18}} (\bibinfo {year} {2020})}\BibitemShut {NoStop}%
\bibitem [{\citenamefont {Alba}\ \emph {et~al.}(2021)\citenamefont {Alba},
  \citenamefont {Bertini}, \citenamefont {Fagotti}, \citenamefont {Piroli},\
  and\ \citenamefont {Ruggiero}}]{alba2021generalized}%
  \BibitemOpen
  \bibfield  {author} {\bibinfo {author} {\bibfnamefont {V.}~\bibnamefont
  {Alba}}, \bibinfo {author} {\bibfnamefont {B.}~\bibnamefont {Bertini}},
  \bibinfo {author} {\bibfnamefont {M.}~\bibnamefont {Fagotti}}, \bibinfo
  {author} {\bibfnamefont {L.}~\bibnamefont {Piroli}},\ and\ \bibinfo {author}
  {\bibfnamefont {P.}~\bibnamefont {Ruggiero}},\ }\bibfield  {title} {\bibinfo
  {title} {Generalized-hydrodynamic approach to inhomogeneous integrable
  systems},\ }\href {https://doi.org/10.21468/SciPostPhys.10.4.094} {\bibfield
  {journal} {\bibinfo  {journal} {SciPost Physics}\ }\textbf {\bibinfo {volume}
  {10}},\ \bibinfo {pages} {094} (\bibinfo {year} {2021})}\BibitemShut
  {NoStop}%
\bibitem [{\citenamefont {Doyon}\ \emph {et~al.}(2025)\citenamefont {Doyon},
  \citenamefont {Gopalakrishnan}, \citenamefont {Moller}, \citenamefont
  {Schmiedmayer},\ and\ \citenamefont {Vasseur}}]{doyon2024generalized}%
  \BibitemOpen
  \bibfield  {author} {\bibinfo {author} {\bibfnamefont {B.}~\bibnamefont
  {Doyon}}, \bibinfo {author} {\bibfnamefont {S.}~\bibnamefont
  {Gopalakrishnan}}, \bibinfo {author} {\bibfnamefont {F.}~\bibnamefont
  {Moller}}, \bibinfo {author} {\bibfnamefont {J.}~\bibnamefont
  {Schmiedmayer}},\ and\ \bibinfo {author} {\bibfnamefont {R.}~\bibnamefont
  {Vasseur}},\ }\bibfield  {title} {\bibinfo {title} {Generalized
  hydrodynamics: A perspective},\ }\href
  {https://doi.org/10.1103/PhysRevX.15.010501} {\bibfield  {journal} {\bibinfo
  {journal} {Phys. Rev. X}\ }\textbf {\bibinfo {volume} {15}},\ \bibinfo
  {pages} {010501} (\bibinfo {year} {2025})}\BibitemShut {NoStop}%
\bibitem [{\citenamefont {Schemmer}\ \emph {et~al.}(2019)\citenamefont
  {Schemmer}, \citenamefont {Bouchoule}, \citenamefont {Doyon},\ and\
  \citenamefont {Dubail}}]{Schemmer2019Generalized}%
  \BibitemOpen
  \bibfield  {author} {\bibinfo {author} {\bibfnamefont {M.}~\bibnamefont
  {Schemmer}}, \bibinfo {author} {\bibfnamefont {I.}~\bibnamefont {Bouchoule}},
  \bibinfo {author} {\bibfnamefont {B.}~\bibnamefont {Doyon}},\ and\ \bibinfo
  {author} {\bibfnamefont {J.}~\bibnamefont {Dubail}},\ }\bibfield  {title}
  {\bibinfo {title} {Generalized hydrodynamics on an atom chip},\ }\href
  {https://doi.org/10.1103/PhysRevLett.122.090601} {\bibfield  {journal}
  {\bibinfo  {journal} {Phys. Rev. Lett.}\ }\textbf {\bibinfo {volume} {122}},\
  \bibinfo {pages} {090601} (\bibinfo {year} {2019})}\BibitemShut {NoStop}%
\bibitem [{\citenamefont {Malvania}\ \emph {et~al.}(2021)\citenamefont
  {Malvania}, \citenamefont {Zhang}, \citenamefont {Le}, \citenamefont
  {Dubail}, \citenamefont {Rigol},\ and\ \citenamefont
  {Weiss}}]{Malvania2021Generalized}%
  \BibitemOpen
  \bibfield  {author} {\bibinfo {author} {\bibfnamefont {N.}~\bibnamefont
  {Malvania}}, \bibinfo {author} {\bibfnamefont {Y.}~\bibnamefont {Zhang}},
  \bibinfo {author} {\bibfnamefont {Y.}~\bibnamefont {Le}}, \bibinfo {author}
  {\bibfnamefont {J.}~\bibnamefont {Dubail}}, \bibinfo {author} {\bibfnamefont
  {M.}~\bibnamefont {Rigol}},\ and\ \bibinfo {author} {\bibfnamefont {D.~S.}\
  \bibnamefont {Weiss}},\ }\bibfield  {title} {\bibinfo {title} {Generalized
  hydrodynamics in strongly interacting 1d bose gases},\ }\href
  {https://doi.org/10.1126/science.abf0147} {\bibfield  {journal} {\bibinfo
  {journal} {Science}\ }\textbf {\bibinfo {volume} {373}},\ \bibinfo {pages}
  {1129} (\bibinfo {year} {2021})}\BibitemShut {NoStop}%
\bibitem [{\citenamefont {Bouchoule}\ and\ \citenamefont
  {Dubail}(2022)}]{Bouchoule2022Generalized}%
  \BibitemOpen
  \bibfield  {author} {\bibinfo {author} {\bibfnamefont {I.}~\bibnamefont
  {Bouchoule}}\ and\ \bibinfo {author} {\bibfnamefont {J.}~\bibnamefont
  {Dubail}},\ }\bibfield  {title} {\bibinfo {title} {Generalized hydrodynamics
  in the one-dimensional bose gas: theory and experiments},\ }\href
  {https://doi.org/10.1088/1742-5468/ac3659} {\bibfield  {journal} {\bibinfo
  {journal} {J. Stat. Mech.}\ }\textbf {\bibinfo {volume} {014003}} (\bibinfo
  {year} {2022})}\BibitemShut {NoStop}%
\bibitem [{\citenamefont {Deutsch}(1991)}]{deutsch1991quantum}%
  \BibitemOpen
  \bibfield  {author} {\bibinfo {author} {\bibfnamefont {J.~M.}\ \bibnamefont
  {Deutsch}},\ }\bibfield  {title} {\bibinfo {title} {Quantum statistical
  mechanics in a closed system},\ }\href
  {https://doi.org/10.1103/physreva.43.2046} {\bibfield  {journal} {\bibinfo
  {journal} {Phys. Rev. A}\ }\textbf {\bibinfo {volume} {43}},\ \bibinfo
  {pages} {2046} (\bibinfo {year} {1991})}\BibitemShut {NoStop}%
\bibitem [{\citenamefont {Srednicki}(1994)}]{srednicki1994chaos}%
  \BibitemOpen
  \bibfield  {author} {\bibinfo {author} {\bibfnamefont {M.}~\bibnamefont
  {Srednicki}},\ }\bibfield  {title} {\bibinfo {title} {Chaos and quantum
  thermalization},\ }\href {https://doi.org/10.1103/physreve.50.888} {\bibfield
   {journal} {\bibinfo  {journal} {Phys. Rev. E}\ }\textbf {\bibinfo {volume}
  {50}},\ \bibinfo {pages} {888} (\bibinfo {year} {1994})}\BibitemShut
  {NoStop}%
\bibitem [{\citenamefont {Srednicki}(1996)}]{srednicki1996thermal}%
  \BibitemOpen
  \bibfield  {author} {\bibinfo {author} {\bibfnamefont {M.}~\bibnamefont
  {Srednicki}},\ }\bibfield  {title} {\bibinfo {title} {Thermal fluctuations in
  quantized chaotic systems},\ }\href
  {https://doi.org/10.1088/0305-4470/29/4/003} {\bibfield  {journal} {\bibinfo
  {journal} {J. Phys. A: Math. Gen.}\ }\textbf {\bibinfo {volume} {29}},\
  \bibinfo {pages} {L75} (\bibinfo {year} {1996})}\BibitemShut {NoStop}%
\bibitem [{\citenamefont {Srednicki}(1999)}]{srednicki1999approach}%
  \BibitemOpen
  \bibfield  {author} {\bibinfo {author} {\bibfnamefont {M.}~\bibnamefont
  {Srednicki}},\ }\bibfield  {title} {\bibinfo {title} {The approach to thermal
  equilibrium in quantized chaotic systems},\ }\href
  {https://doi.org/10.1088/0305-4470/32/7/007} {\bibfield  {journal} {\bibinfo
  {journal} {J. Phys. A: Math. Gen.}\ }\textbf {\bibinfo {volume} {32}},\
  \bibinfo {pages} {1163} (\bibinfo {year} {1999})}\BibitemShut {NoStop}%
\bibitem [{\citenamefont {Rigol}\ \emph {et~al.}(2008)\citenamefont {Rigol},
  \citenamefont {Dunjko},\ and\ \citenamefont
  {Olshanii}}]{rigol2008thermalization}%
  \BibitemOpen
  \bibfield  {author} {\bibinfo {author} {\bibfnamefont {M.}~\bibnamefont
  {Rigol}}, \bibinfo {author} {\bibfnamefont {V.}~\bibnamefont {Dunjko}},\ and\
  \bibinfo {author} {\bibfnamefont {M.}~\bibnamefont {Olshanii}},\ }\bibfield
  {title} {\bibinfo {title} {Thermalization and its mechanism for generic
  isolated quantum systems},\ }\href {https://doi.org/10.1038/nature06838}
  {\bibfield  {journal} {\bibinfo  {journal} {Nature}\ }\textbf {\bibinfo
  {volume} {452}},\ \bibinfo {pages} {854} (\bibinfo {year}
  {2008})}\BibitemShut {NoStop}%
\bibitem [{\citenamefont {Jansen}\ \emph {et~al.}(2019)\citenamefont {Jansen},
  \citenamefont {Stolpp}, \citenamefont {Vidmar},\ and\ \citenamefont
  {Heidrich-Meisner}}]{eth-num-PhysRevB.99.155130}%
  \BibitemOpen
  \bibfield  {author} {\bibinfo {author} {\bibfnamefont {D.}~\bibnamefont
  {Jansen}}, \bibinfo {author} {\bibfnamefont {J.}~\bibnamefont {Stolpp}},
  \bibinfo {author} {\bibfnamefont {L.}~\bibnamefont {Vidmar}},\ and\ \bibinfo
  {author} {\bibfnamefont {F.}~\bibnamefont {Heidrich-Meisner}},\ }\bibfield
  {title} {\bibinfo {title} {Eigenstate thermalization and quantum chaos in the
  holstein polaron model},\ }\href {https://doi.org/10.1103/PhysRevB.99.155130}
  {\bibfield  {journal} {\bibinfo  {journal} {Phys. Rev. B}\ }\textbf {\bibinfo
  {volume} {99}},\ \bibinfo {pages} {155130} (\bibinfo {year}
  {2019})}\BibitemShut {NoStop}%
\bibitem [{\citenamefont {LeBlond}\ \emph {et~al.}(2019)\citenamefont
  {LeBlond}, \citenamefont {Mallayya}, \citenamefont {Vidmar},\ and\
  \citenamefont {Rigol}}]{eth-num-PhysRevE.100.062134}%
  \BibitemOpen
  \bibfield  {author} {\bibinfo {author} {\bibfnamefont {T.}~\bibnamefont
  {LeBlond}}, \bibinfo {author} {\bibfnamefont {K.}~\bibnamefont {Mallayya}},
  \bibinfo {author} {\bibfnamefont {L.}~\bibnamefont {Vidmar}},\ and\ \bibinfo
  {author} {\bibfnamefont {M.}~\bibnamefont {Rigol}},\ }\bibfield  {title}
  {\bibinfo {title} {Entanglement and matrix elements of observables in
  interacting integrable systems},\ }\href
  {https://doi.org/10.1103/PhysRevE.100.062134} {\bibfield  {journal} {\bibinfo
   {journal} {Phys. Rev. E}\ }\textbf {\bibinfo {volume} {100}},\ \bibinfo
  {pages} {062134} (\bibinfo {year} {2019})}\BibitemShut {NoStop}%
\bibitem [{\citenamefont {Santos}\ and\ \citenamefont
  {Rigol}(2010)}]{eth-num-PhysRevE.82.031130}%
  \BibitemOpen
  \bibfield  {author} {\bibinfo {author} {\bibfnamefont {L.~F.}\ \bibnamefont
  {Santos}}\ and\ \bibinfo {author} {\bibfnamefont {M.}~\bibnamefont {Rigol}},\
  }\bibfield  {title} {\bibinfo {title} {Localization and the effects of
  symmetries in the thermalization properties of one-dimensional quantum
  systems},\ }\href {https://doi.org/10.1103/PhysRevE.82.031130} {\bibfield
  {journal} {\bibinfo  {journal} {Phys. Rev. E}\ }\textbf {\bibinfo {volume}
  {82}},\ \bibinfo {pages} {031130} (\bibinfo {year} {2010})}\BibitemShut
  {NoStop}%
\bibitem [{\citenamefont {Steinigeweg}\ \emph {et~al.}(2013)\citenamefont
  {Steinigeweg}, \citenamefont {Herbrych},\ and\ \citenamefont
  {Prelov\ifmmode~\check{s}\else \v{s}\fi{}ek}}]{eth-num-PhysRevE.87.012118}%
  \BibitemOpen
  \bibfield  {author} {\bibinfo {author} {\bibfnamefont {R.}~\bibnamefont
  {Steinigeweg}}, \bibinfo {author} {\bibfnamefont {J.}~\bibnamefont
  {Herbrych}},\ and\ \bibinfo {author} {\bibfnamefont {P.}~\bibnamefont
  {Prelov\ifmmode~\check{s}\else \v{s}\fi{}ek}},\ }\bibfield  {title} {\bibinfo
  {title} {Eigenstate thermalization within isolated spin-chain systems},\
  }\href {https://doi.org/10.1103/PhysRevE.87.012118} {\bibfield  {journal}
  {\bibinfo  {journal} {Phys. Rev. E}\ }\textbf {\bibinfo {volume} {87}},\
  \bibinfo {pages} {012118} (\bibinfo {year} {2013})}\BibitemShut {NoStop}%
\bibitem [{\citenamefont {Beugeling}\ \emph {et~al.}(2014)\citenamefont
  {Beugeling}, \citenamefont {Moessner},\ and\ \citenamefont
  {Haque}}]{eth-num-PhysRevE.89.042112}%
  \BibitemOpen
  \bibfield  {author} {\bibinfo {author} {\bibfnamefont {W.}~\bibnamefont
  {Beugeling}}, \bibinfo {author} {\bibfnamefont {R.}~\bibnamefont
  {Moessner}},\ and\ \bibinfo {author} {\bibfnamefont {M.}~\bibnamefont
  {Haque}},\ }\bibfield  {title} {\bibinfo {title} {Finite-size scaling of
  eigenstate thermalization},\ }\href
  {https://doi.org/10.1103/PhysRevE.89.042112} {\bibfield  {journal} {\bibinfo
  {journal} {Phys. Rev. E}\ }\textbf {\bibinfo {volume} {89}},\ \bibinfo
  {pages} {042112} (\bibinfo {year} {2014})}\BibitemShut {NoStop}%
\bibitem [{\citenamefont {Kim}\ \emph {et~al.}(2014)\citenamefont {Kim},
  \citenamefont {Ikeda},\ and\ \citenamefont
  {Huse}}]{eth-num-PhysRevE.90.052105}%
  \BibitemOpen
  \bibfield  {author} {\bibinfo {author} {\bibfnamefont {H.}~\bibnamefont
  {Kim}}, \bibinfo {author} {\bibfnamefont {T.~N.}\ \bibnamefont {Ikeda}},\
  and\ \bibinfo {author} {\bibfnamefont {D.~A.}\ \bibnamefont {Huse}},\
  }\bibfield  {title} {\bibinfo {title} {Testing whether all eigenstates obey
  the eigenstate thermalization hypothesis},\ }\href
  {https://doi.org/10.1103/PhysRevE.90.052105} {\bibfield  {journal} {\bibinfo
  {journal} {Phys. Rev. E}\ }\textbf {\bibinfo {volume} {90}},\ \bibinfo
  {pages} {052105} (\bibinfo {year} {2014})}\BibitemShut {NoStop}%
\bibitem [{\citenamefont {Beugeling}\ \emph {et~al.}(2015)\citenamefont
  {Beugeling}, \citenamefont {Moessner},\ and\ \citenamefont
  {Haque}}]{eth-num-PhysRevE.91.012144}%
  \BibitemOpen
  \bibfield  {author} {\bibinfo {author} {\bibfnamefont {W.}~\bibnamefont
  {Beugeling}}, \bibinfo {author} {\bibfnamefont {R.}~\bibnamefont
  {Moessner}},\ and\ \bibinfo {author} {\bibfnamefont {M.}~\bibnamefont
  {Haque}},\ }\bibfield  {title} {\bibinfo {title} {Off-diagonal matrix
  elements of local operators in many-body quantum systems},\ }\href
  {https://doi.org/10.1103/PhysRevE.91.012144} {\bibfield  {journal} {\bibinfo
  {journal} {Phys. Rev. E}\ }\textbf {\bibinfo {volume} {91}},\ \bibinfo
  {pages} {012144} (\bibinfo {year} {2015})}\BibitemShut {NoStop}%
\bibitem [{\citenamefont {Mondaini}\ \emph {et~al.}(2016)\citenamefont
  {Mondaini}, \citenamefont {Fratus}, \citenamefont {Srednicki},\ and\
  \citenamefont {Rigol}}]{eth-num-PhysRevE.93.032104}%
  \BibitemOpen
  \bibfield  {author} {\bibinfo {author} {\bibfnamefont {R.}~\bibnamefont
  {Mondaini}}, \bibinfo {author} {\bibfnamefont {K.~R.}\ \bibnamefont
  {Fratus}}, \bibinfo {author} {\bibfnamefont {M.}~\bibnamefont {Srednicki}},\
  and\ \bibinfo {author} {\bibfnamefont {M.}~\bibnamefont {Rigol}},\ }\bibfield
   {title} {\bibinfo {title} {Eigenstate thermalization in the two-dimensional
  transverse field ising model},\ }\href
  {https://doi.org/10.1103/PhysRevE.93.032104} {\bibfield  {journal} {\bibinfo
  {journal} {Phys. Rev. E}\ }\textbf {\bibinfo {volume} {93}},\ \bibinfo
  {pages} {032104} (\bibinfo {year} {2016})}\BibitemShut {NoStop}%
\bibitem [{\citenamefont {Mondaini}\ and\ \citenamefont
  {Rigol}(2017)}]{eth-num-PhysRevE.96.012157}%
  \BibitemOpen
  \bibfield  {author} {\bibinfo {author} {\bibfnamefont {R.}~\bibnamefont
  {Mondaini}}\ and\ \bibinfo {author} {\bibfnamefont {M.}~\bibnamefont
  {Rigol}},\ }\bibfield  {title} {\bibinfo {title} {Eigenstate thermalization
  in the two-dimensional transverse field ising model. ii. off-diagonal matrix
  elements of observables},\ }\href
  {https://doi.org/10.1103/PhysRevE.96.012157} {\bibfield  {journal} {\bibinfo
  {journal} {Phys. Rev. E}\ }\textbf {\bibinfo {volume} {96}},\ \bibinfo
  {pages} {012157} (\bibinfo {year} {2017})}\BibitemShut {NoStop}%
\bibitem [{\citenamefont {Delacretaz}\ \emph {et~al.}()\citenamefont
  {Delacretaz}, \citenamefont {Fitzpatrick}, \citenamefont {Katz},\ and\
  \citenamefont {Walters}}]{Delacretaz:2022ojg}%
  \BibitemOpen
  \bibfield  {author} {\bibinfo {author} {\bibfnamefont {L.~V.}\ \bibnamefont
  {Delacretaz}}, \bibinfo {author} {\bibfnamefont {A.~L.}\ \bibnamefont
  {Fitzpatrick}}, \bibinfo {author} {\bibfnamefont {E.}~\bibnamefont {Katz}},\
  and\ \bibinfo {author} {\bibfnamefont {M.~T.}\ \bibnamefont {Walters}},\
  }\bibfield  {title} {\bibinfo {title} {Thermalization and chaos in a 1+1d
  QFT},\ }\href {https://doi.org/10.1007/JHEP02(2023)045} {\bibfield  {journal}
  {\bibinfo  {journal} {J. High Energy Phys.}} {\bibinfo {volume}
  {02}}\bibinfo  {number} { (2023)}\ \bibinfo {pages} {045}}\BibitemShut
  {NoStop}%
\bibitem [{\citenamefont {Foini}\ and\ \citenamefont
  {Kurchan}(2019)}]{foini2019eigenstate}%
  \BibitemOpen
\bibfield  {number} {  }\bibfield  {author} {\bibinfo {author} {\bibfnamefont
  {L.}~\bibnamefont {Foini}}\ and\ \bibinfo {author} {\bibfnamefont
  {J.}~\bibnamefont {Kurchan}},\ }\bibfield  {title} {\bibinfo {title}
  {Eigenstate thermalization hypothesis and out of time order correlators},\
  }\href {https://doi.org/10.1103/PhysRevE.99.042139} {\bibfield  {journal}
  {\bibinfo  {journal} {Phys. Rev. E}\ }\textbf {\bibinfo {volume} {99}},\
  \bibinfo {pages} {042139} (\bibinfo {year} {2019})}\BibitemShut {NoStop}%
\bibitem [{\citenamefont {Chan}\ \emph {et~al.}(2019)\citenamefont {Chan},
  \citenamefont {De~Luca},\ and\ \citenamefont {Chalker}}]{chan2019eigenstate}%
  \BibitemOpen
  \bibfield  {author} {\bibinfo {author} {\bibfnamefont {A.}~\bibnamefont
  {Chan}}, \bibinfo {author} {\bibfnamefont {A.}~\bibnamefont {De~Luca}},\ and\
  \bibinfo {author} {\bibfnamefont {J.~T.}\ \bibnamefont {Chalker}},\
  }\bibfield  {title} {\bibinfo {title} {Eigenstate correlations,
  thermalization, and the butterfly effect},\ }\href
  {https://doi.org/10.1103/PhysRevLett.122.220601} {\bibfield  {journal}
  {\bibinfo  {journal} {Phys. Rev. Lett.}\ }\textbf {\bibinfo {volume} {122}},\
  \bibinfo {pages} {220601} (\bibinfo {year} {2019})}\BibitemShut {NoStop}%
\bibitem [{\citenamefont {Murthy}\ and\ \citenamefont
  {Srednicki}(2019)}]{murthy2019bounds}%
  \BibitemOpen
  \bibfield  {author} {\bibinfo {author} {\bibfnamefont {C.}~\bibnamefont
  {Murthy}}\ and\ \bibinfo {author} {\bibfnamefont {M.}~\bibnamefont
  {Srednicki}},\ }\bibfield  {title} {\bibinfo {title} {Bounds on chaos from
  the eigenstate thermalization hypothesis},\ }\href
  {https://doi.org/10.1103/PhysRevLett.123.230606} {\bibfield  {journal}
  {\bibinfo  {journal} {Phys. Rev. Lett.}\ }\textbf {\bibinfo {volume} {123}},\
  \bibinfo {pages} {230606} (\bibinfo {year} {2019})}\BibitemShut {NoStop}%
\bibitem [{\citenamefont {Richter}\ \emph
  {et~al.}(2020{\natexlab{a}})\citenamefont {Richter}, \citenamefont
  {Dymarsky}, \citenamefont {Steinigeweg},\ and\ \citenamefont
  {Gemmer}}]{richter2020eigenstate}%
  \BibitemOpen
  \bibfield  {author} {\bibinfo {author} {\bibfnamefont {J.}~\bibnamefont
  {Richter}}, \bibinfo {author} {\bibfnamefont {A.}~\bibnamefont {Dymarsky}},
  \bibinfo {author} {\bibfnamefont {R.}~\bibnamefont {Steinigeweg}},\ and\
  \bibinfo {author} {\bibfnamefont {J.}~\bibnamefont {Gemmer}},\ }\bibfield
  {title} {\bibinfo {title} {Eigenstate thermalization hypothesis beyond
  standard indicators: Emergence of random-matrix behavior at small
  frequencies},\ }\href {https://doi.org/10.1103/PhysRevE.102.042127}
  {\bibfield  {journal} {\bibinfo  {journal} {Phys. Rev. E}\ }\textbf {\bibinfo
  {volume} {102}},\ \bibinfo {pages} {042127} (\bibinfo {year}
  {2020}{\natexlab{a}})}\BibitemShut {NoStop}%
\bibitem [{\citenamefont {Brenes}\ \emph {et~al.}(2021)\citenamefont {Brenes},
  \citenamefont {Pappalardi}, \citenamefont {Mitchison}, \citenamefont
  {Goold},\ and\ \citenamefont {Silva}}]{brenes2021out}%
  \BibitemOpen
  \bibfield  {author} {\bibinfo {author} {\bibfnamefont {M.}~\bibnamefont
  {Brenes}}, \bibinfo {author} {\bibfnamefont {S.}~\bibnamefont {Pappalardi}},
  \bibinfo {author} {\bibfnamefont {M.~T.}\ \bibnamefont {Mitchison}}, \bibinfo
  {author} {\bibfnamefont {J.}~\bibnamefont {Goold}},\ and\ \bibinfo {author}
  {\bibfnamefont {A.}~\bibnamefont {Silva}},\ }\bibfield  {title} {\bibinfo
  {title} {Out-of-time-order correlations and the fine structure of eigenstate
  thermalization},\ }\href {https://doi.org/10.1103/PhysRevE.104.034120}
  {\bibfield  {journal} {\bibinfo  {journal} {Phys. Rev. E}\ }\textbf {\bibinfo
  {volume} {104}},\ \bibinfo {pages} {034120} (\bibinfo {year}
  {2021})}\BibitemShut {NoStop}%
\bibitem [{\citenamefont {Wang}\ \emph {et~al.}(2022)\citenamefont {Wang},
  \citenamefont {Lamann}, \citenamefont {Richter}, \citenamefont {Steinigeweg},
  \citenamefont {Dymarsky},\ and\ \citenamefont {Gemmer}}]{wang2021eigenstate}%
  \BibitemOpen
  \bibfield  {author} {\bibinfo {author} {\bibfnamefont {J.}~\bibnamefont
  {Wang}}, \bibinfo {author} {\bibfnamefont {M.~H.}\ \bibnamefont {Lamann}},
  \bibinfo {author} {\bibfnamefont {J.}~\bibnamefont {Richter}}, \bibinfo
  {author} {\bibfnamefont {R.}~\bibnamefont {Steinigeweg}}, \bibinfo {author}
  {\bibfnamefont {A.}~\bibnamefont {Dymarsky}},\ and\ \bibinfo {author}
  {\bibfnamefont {J.}~\bibnamefont {Gemmer}},\ }\bibfield  {title} {\bibinfo
  {title} {Eigenstate thermalization hypothesis and its deviations from
  random-matrix theory beyond the thermalization time},\ }\href
  {https://doi.org/10.1103/PhysRevLett.128.180601} {\bibfield  {journal}
  {\bibinfo  {journal} {Phys. Rev. Lett.}\ }\textbf {\bibinfo {volume} {128}},\
  \bibinfo {pages} {180601} (\bibinfo {year} {2022})}\BibitemShut {NoStop}%
\bibitem [{\citenamefont {Dymarsky}(2022)}]{dymarsky2022bound}%
  \BibitemOpen
  \bibfield  {author} {\bibinfo {author} {\bibfnamefont {A.}~\bibnamefont
  {Dymarsky}},\ }\bibfield  {title} {\bibinfo {title} {Bound on eigenstate
  thermalization from transport},\ }\href
  {https://doi.org/10.1103/PhysRevLett.128.190601} {\bibfield  {journal}
  {\bibinfo  {journal} {Phys. Rev. Lett.}\ }\textbf {\bibinfo {volume} {128}},\
  \bibinfo {pages} {190601} (\bibinfo {year} {2022})}\BibitemShut {NoStop}%
\bibitem [{foo({\natexlab{a}})}]{footnote2}%
  \BibitemOpen
  \href@noop {} {}\bibinfo {note} {For $n=1,2$, one
  recovers the standard ETH \cite{srednicki1999approach}, where $F_e^{(1)} =
  {\cal A}(e)$ is the constant equilibrium average and $F_e^{(2)}(\omega)=|f(e,
  \omega)|^2$ is the dynamical correlation function.}\BibitemShut {Stop}%
\bibitem [{\citenamefont {Pappalardi}\ \emph {et~al.}(2024)\citenamefont
  {Pappalardi}, \citenamefont {Foini},\ and\ \citenamefont
  {Kurchan}}]{pappalardi2024microcanonical}%
  \BibitemOpen
  \bibfield  {author} {\bibinfo {author} {\bibfnamefont {S.}~\bibnamefont
  {Pappalardi}}, \bibinfo {author} {\bibfnamefont {L.}~\bibnamefont {Foini}},\
  and\ \bibinfo {author} {\bibfnamefont {J.}~\bibnamefont {Kurchan}},\
  }\bibfield  {title} {\bibinfo {title} {Microcanonical windows on quantum
  operators},\ }\href {https://doi.org/10.22331/q-2024-01-11-1227} {\bibfield
  {journal} {\bibinfo  {journal} {Quantum}\ }\textbf {\bibinfo {volume} {8}},\
  \bibinfo {pages} {1227} (\bibinfo {year} {2024})}\BibitemShut {NoStop}%
\bibitem [{\citenamefont {Pappalardi}\ \emph {et~al.}(2022)\citenamefont
  {Pappalardi}, \citenamefont {Foini},\ and\ \citenamefont
  {Kurchan}}]{pappalardi2022eigenstate}%
  \BibitemOpen
  \bibfield  {author} {\bibinfo {author} {\bibfnamefont {S.}~\bibnamefont
  {Pappalardi}}, \bibinfo {author} {\bibfnamefont {L.}~\bibnamefont {Foini}},\
  and\ \bibinfo {author} {\bibfnamefont {J.}~\bibnamefont {Kurchan}},\
  }\bibfield  {title} {\bibinfo {title} {Eigenstate thermalization hypothesis
  and free probability},\ }\href
  {https://doi.org/10.1103/PhysRevLett.129.170603} {\bibfield  {journal}
  {\bibinfo  {journal} {Phys. Rev. Lett.}\ }\textbf {\bibinfo {volume} {129}},\
  \bibinfo {pages} {170603} (\bibinfo {year} {2022})}\BibitemShut {NoStop}%
\bibitem [{\citenamefont {Voiculescu}(1985)}]{Voiculescu1985Symmetries}%
  \BibitemOpen
  \bibfield  {author} {\bibinfo {author} {\bibfnamefont {D.}~\bibnamefont
  {Voiculescu}},\ }\bibfield  {title} {\bibinfo {title} {Symmetries of some
  reduced free product $c^*$-algebras},\ }in\ \href
  {https://doi.org/10.1007/BFb0074909} {\emph {\bibinfo {booktitle} {Operator
  Algebras and Their Connections with Topology and Ergodic Theory}}},\ \bibinfo
  {series} {Lecture Notes in Mathematics}, Vol.\ \bibinfo {volume} {1132}\
  (\bibinfo  {publisher} {Springer},\ \bibinfo {year} {1985})\ pp.\ \bibinfo
  {pages} {556--588}\BibitemShut {NoStop}%
\bibitem [{\citenamefont {Mingo}\ and\ \citenamefont
  {Speicher}(2017)}]{Mingo2017Free}%
  \BibitemOpen
  \bibfield  {author} {\bibinfo {author} {\bibfnamefont {J.~A.}\ \bibnamefont
  {Mingo}}\ and\ \bibinfo {author} {\bibfnamefont {R.}~\bibnamefont
  {Speicher}},\ }\href
  {https://link.springer.com/book/10.1007/978-1-4939-6942-5} {\emph {\bibinfo
  {title} {Free probability and random matrices}}},\ Vol.~\bibinfo {volume}
  {35}\ (\bibinfo  {publisher} {Springer},\ \bibinfo {year} {2017})\BibitemShut
  {NoStop}%
\bibitem [{\citenamefont {Speicher}(1997)}]{Speicher1997Free}%
  \BibitemOpen
  \bibfield  {author} {\bibinfo {author} {\bibfnamefont {R.}~\bibnamefont
  {Speicher}},\ }\bibfield  {title} {\bibinfo {title} {Free probability theory
  and non-crossing partitions.},\ }\href
  {https://citeseerx.ist.psu.edu/viewdoc/download?doi=10.1.1.34.6530&rep=rep1&type=pdf}
  {\bibfield  {journal} {\bibinfo  {journal} {S{\'e}minaire Lotharingien de
  Combinatoire [electronic only]}\ }\textbf {\bibinfo {volume} {39}},\ \bibinfo
  {pages} {B39c} (\bibinfo {year} {1997})}\BibitemShut {NoStop}%
\bibitem [{\citenamefont {Pappalardi}\ \emph {et~al.}(2025)\citenamefont
  {Pappalardi}, \citenamefont {Fritzsch},\ and\ \citenamefont
  {Prosen}}]{Pappalardi2025Full}%
  \BibitemOpen
  \bibfield  {author} {\bibinfo {author} {\bibfnamefont {S.}~\bibnamefont
  {Pappalardi}}, \bibinfo {author} {\bibfnamefont {F.}~\bibnamefont
  {Fritzsch}},\ and\ \bibinfo {author} {\bibfnamefont {T.}~\bibnamefont
  {Prosen}},\ }\bibfield  {title} {\bibinfo {title} {Full eigenstate
  thermalization via free cumulants in quantum lattice systems},\ }\href
  {https://doi.org/10.1103/PhysRevLett.134.140404} {\bibfield  {journal}
  {\bibinfo  {journal} {Phys. Rev. Lett.}\ }\textbf {\bibinfo {volume} {134}},\
  \bibinfo {pages} {140404} (\bibinfo {year} {2025})}\BibitemShut {NoStop}%
\bibitem [{\citenamefont {Wang}\ \emph {et~al.}(2024)\citenamefont {Wang},
  \citenamefont {Richter}, \citenamefont {Lamann}, \citenamefont {Steinigeweg},
  \citenamefont {Gemmer},\ and\ \citenamefont {Dymarsky}}]{wang2024emergence}%
  \BibitemOpen
  \bibfield  {author} {\bibinfo {author} {\bibfnamefont {J.}~\bibnamefont
  {Wang}}, \bibinfo {author} {\bibfnamefont {J.}~\bibnamefont {Richter}},
  \bibinfo {author} {\bibfnamefont {M.~H.}\ \bibnamefont {Lamann}}, \bibinfo
  {author} {\bibfnamefont {R.}~\bibnamefont {Steinigeweg}}, \bibinfo {author}
  {\bibfnamefont {J.}~\bibnamefont {Gemmer}},\ and\ \bibinfo {author}
  {\bibfnamefont {A.}~\bibnamefont {Dymarsky}},\ }\bibfield  {title} {\bibinfo
  {title} {Emergence of unitary symmetry of microcanonically truncated
  operators in chaotic quantum systems},\ }\href
  {https://doi.org/10.1103/PhysRevE.110.L032203} {\bibfield  {journal}
  {\bibinfo  {journal} {Phys. Rev. E}\ }\textbf {\bibinfo {volume} {110}},\
  \bibinfo {pages} {L032203} (\bibinfo {year} {2024})}\BibitemShut {NoStop}%
\bibitem [{\citenamefont {Jindal}\ and\ \citenamefont
  {Hosur}()}]{Jindal2024Generalized}%
  \BibitemOpen
  \bibfield  {author} {\bibinfo {author} {\bibfnamefont {S.}~\bibnamefont
  {Jindal}}\ and\ \bibinfo {author} {\bibfnamefont {P.}~\bibnamefont {Hosur}},\
  }\bibfield  {title} {\bibinfo {title} {Generalized free cumulants for quantum
  chaotic systems},\ }\href {https://doi.org/10.1007/JHEP01(2024)066}
  {\bibfield  {journal} {\bibinfo  {journal} {J. High Energy Phys.}}
  {\bibinfo {volume} {01}}\bibinfo  {number} { (2024)}\ \bibinfo {pages}
  {066}}\BibitemShut {NoStop}%
\bibitem [{\citenamefont {Fava}\ \emph {et~al.}(2025)\citenamefont {Fava},
  \citenamefont {Kurchan},\ and\ \citenamefont {Pappalardi}}]{Fava2025Designs}%
  \BibitemOpen
\bibfield  {number} {  }\bibfield  {author} {\bibinfo {author} {\bibfnamefont
  {M.}~\bibnamefont {Fava}}, \bibinfo {author} {\bibfnamefont {J.}~\bibnamefont
  {Kurchan}},\ and\ \bibinfo {author} {\bibfnamefont {S.}~\bibnamefont
  {Pappalardi}},\ }\bibfield  {title} {\bibinfo {title} {Designs via free
  probability},\ }\href {https://doi.org/10.1103/PhysRevX.15.011031} {\bibfield
   {journal} {\bibinfo  {journal} {Phys. Rev. X}\ }\textbf {\bibinfo {volume}
  {15}},\ \bibinfo {pages} {011031} (\bibinfo {year} {2025})}\BibitemShut
  {NoStop}%
\bibitem [{\citenamefont {Fritzsch}\ \emph {et~al.}(2025)\citenamefont
  {Fritzsch}, \citenamefont {Prosen},\ and\ \citenamefont
  {Pappalardi}}]{Fritzsch2025Microcanonical}%
  \BibitemOpen
  \bibfield  {author} {\bibinfo {author} {\bibfnamefont {F.}~\bibnamefont
  {Fritzsch}}, \bibinfo {author} {\bibfnamefont {T.}~\bibnamefont {Prosen}},\
  and\ \bibinfo {author} {\bibfnamefont {S.}~\bibnamefont {Pappalardi}},\
  }\bibfield  {title} {\bibinfo {title} {Microcanonical free cumulants in
  lattice systems},\ }\href {https://doi.org/10.1103/PhysRevB.111.054303}
  {\bibfield  {journal} {\bibinfo  {journal} {Phys. Rev. B}\ }\textbf {\bibinfo
  {volume} {111}},\ \bibinfo {pages} {054303} (\bibinfo {year}
  {2025})}\BibitemShut {NoStop}%
\bibitem [{\citenamefont {Chen}\ and\ \citenamefont
  {Kudler-Flam}(2025)}]{Chen2025FreeIndependence}%
  \BibitemOpen
  \bibfield  {author} {\bibinfo {author} {\bibfnamefont {H.~J.}\ \bibnamefont
  {Chen}}\ and\ \bibinfo {author} {\bibfnamefont {J.}~\bibnamefont
  {Kudler-Flam}},\ }\bibfield  {title} {\bibinfo {title} {Free independence and
  the noncrossing partition lattice in dual-unitary quantum circuits},\ }\href
  {https://doi.org/10.1103/PhysRevB.111.014311} {\bibfield  {journal} {\bibinfo
   {journal} {Phys. Rev. B}\ }\textbf {\bibinfo {volume} {111}},\ \bibinfo
  {pages} {014311} (\bibinfo {year} {2025})}\BibitemShut {NoStop}%
\bibitem [{\citenamefont {Vallini}\ and\ \citenamefont
  {Pappalardi}(2024)}]{Vallini2024LongTime}%
  \BibitemOpen
  \bibfield  {author} {\bibinfo {author} {\bibfnamefont {E.}~\bibnamefont
  {Vallini}}\ and\ \bibinfo {author} {\bibfnamefont {S.}~\bibnamefont
  {Pappalardi}},\ }\bibfield  {title} {\bibinfo {title} {Long-time freeness in
  the kicked top},\ }\href {https://arxiv.org/abs/2411.12050} {\bibfield
  {journal} {\bibinfo  {journal} {arXiv preprint arXiv:2411.12050}\ } (\bibinfo
  {year} {2024})}\BibitemShut {NoStop}%
\bibitem [{\citenamefont {Ampelogiannis}\ and\ \citenamefont
  {Doyon}(2025)}]{ampelogiannis2024cluster}%
  \BibitemOpen
  \bibfield  {author} {\bibinfo {author} {\bibfnamefont {D.}~\bibnamefont
  {Ampelogiannis}}\ and\ \bibinfo {author} {\bibfnamefont {B.}~\bibnamefont
  {Doyon}},\ }\bibfield  {title} {\bibinfo {title} {Clustering of higher order
  connected correlations in {C*} dynamical systems},\ }\href
  {https://doi.org/10.1063/5.023361} {\bibfield  {journal} {\bibinfo  {journal}
  {J. Math. Phys.}\ }\textbf {\bibinfo {volume} {66}},\ \bibinfo {pages}
  {053507} (\bibinfo {year} {2025})}\BibitemShut {NoStop}%
\bibitem [{\citenamefont {Camargo}\ \emph {et~al.}(2025)\citenamefont
  {Camargo}, \citenamefont {Fu}, \citenamefont {Jahnke}, \citenamefont {Pal},\
  and\ \citenamefont {Kim}}]{Camargo2025QuantumSignatures}%
  \BibitemOpen
  \bibfield  {author} {\bibinfo {author} {\bibfnamefont {H.~A.}\ \bibnamefont
  {Camargo}}, \bibinfo {author} {\bibfnamefont {Y.}~\bibnamefont {Fu}},
  \bibinfo {author} {\bibfnamefont {V.}~\bibnamefont {Jahnke}}, \bibinfo
  {author} {\bibfnamefont {K.}~\bibnamefont {Pal}},\ and\ \bibinfo {author}
  {\bibfnamefont {K.-Y.}\ \bibnamefont {Kim}},\ }\bibfield  {title} {\bibinfo
  {title} {Quantum signatures of chaos from free probability},\ }\href
  {https://arxiv.org/abs/2503.20338} {\bibfield  {journal} {\bibinfo  {journal}
  {arXiv preprint arXiv:2503.20338}\ } (\bibinfo {year} {2025})}\BibitemShut
  {NoStop}%
\bibitem [{\citenamefont {Alves}\ \emph {et~al.}(2025)\citenamefont {Alves},
  \citenamefont {Fritzsch},\ and\ \citenamefont {Claeys}}]{alves2025probes}%
  \BibitemOpen
  \bibfield  {author} {\bibinfo {author} {\bibfnamefont {G.~O.}\ \bibnamefont
  {Alves}}, \bibinfo {author} {\bibfnamefont {F.}~\bibnamefont {Fritzsch}},\
  and\ \bibinfo {author} {\bibfnamefont {P.~W.}\ \bibnamefont {Claeys}},\
  }\bibfield  {title} {\bibinfo {title} {Probes of full eigenstate
  thermalization in ergodicity-breaking quantum circuits},\ }\href
  {https://doi.org/10.22331/q-2025-12-15-1949} {\bibfield  {journal} {\bibinfo
  {journal} {Quantum}\ }\textbf {\bibinfo {volume} {9}},\ \bibinfo {pages}
  {1949} (\bibinfo {year} {2025})}\BibitemShut {NoStop}%
\bibitem [{\citenamefont {Delacretaz}(2020)}]{Delacretaz:2020nit}%
  \BibitemOpen
  \bibfield  {author} {\bibinfo {author} {\bibfnamefont {L.~V.}\ \bibnamefont
  {Delacretaz}},\ }\bibfield  {title} {\bibinfo {title} {{Heavy operators and
  hydrodynamic tails}},\ }\href {https://doi.org/10.21468/SciPostPhys.9.3.034}
  {\bibfield  {journal} {\bibinfo  {journal} {SciPost Phys.}\ }\textbf
  {\bibinfo {volume} {9}},\ \bibinfo {pages} {034} (\bibinfo {year}
  {2020})}\BibitemShut {NoStop}%
\bibitem [{\citenamefont {Capizzi}\ \emph {et~al.}(2025)\citenamefont
  {Capizzi}, \citenamefont {Wang}, \citenamefont {Xu}, \citenamefont {Mazza},\
  and\ \citenamefont {Poletti}}]{capizzi2024hydrodynamics}%
  \BibitemOpen
  \bibfield  {author} {\bibinfo {author} {\bibfnamefont {L.}~\bibnamefont
  {Capizzi}}, \bibinfo {author} {\bibfnamefont {J.}~\bibnamefont {Wang}},
  \bibinfo {author} {\bibfnamefont {X.}~\bibnamefont {Xu}}, \bibinfo {author}
  {\bibfnamefont {L.}~\bibnamefont {Mazza}},\ and\ \bibinfo {author}
  {\bibfnamefont {D.}~\bibnamefont {Poletti}},\ }\bibfield  {title} {\bibinfo
  {title} {Hydrodynamics and the eigenstate thermalization hypothesis},\ }\href
  {https://link.aps.org/doi/10.1103/PhysRevX.15.011059} {\bibfield  {journal}
  {\bibinfo  {journal} {Phys. Rev. X}\ }\textbf {\bibinfo {volume} {15}},\
  \bibinfo {pages} {011059} (\bibinfo {year} {2025})}\BibitemShut {NoStop}%
\bibitem [{\citenamefont {Delacr{\'e}taz}\ and\ \citenamefont
  {Mishra}(2024)}]{delacretaz2024nonlinear}%
  \BibitemOpen
  \bibfield  {author} {\bibinfo {author} {\bibfnamefont {L.~V.}\ \bibnamefont
  {Delacr{\'e}taz}}\ and\ \bibinfo {author} {\bibfnamefont {R.}~\bibnamefont
  {Mishra}},\ }\bibfield  {title} {\bibinfo {title} {Nonlinear response in
  diffusive systems},\ }\href {https://doi.org/10.21468/SciPostPhys.16.2.047}
  {\bibfield  {journal} {\bibinfo  {journal} {SciPost Physics}\ }\textbf
  {\bibinfo {volume} {16}},\ \bibinfo {pages} {047} (\bibinfo {year}
  {2024})}\BibitemShut {NoStop}%
\bibitem [{\citenamefont {Bartsch}\ and\ \citenamefont
  {Gemmer}(2009)}]{DQT-Gemmer}%
  \BibitemOpen
  \bibfield  {author} {\bibinfo {author} {\bibfnamefont {C.}~\bibnamefont
  {Bartsch}}\ and\ \bibinfo {author} {\bibfnamefont {J.}~\bibnamefont
  {Gemmer}},\ }\bibfield  {title} {\bibinfo {title} {Dynamical typicality of
  quantum expectation values},\ }\href
  {https://doi.org/10.1103/PhysRevLett.102.110403} {\bibfield  {journal}
  {\bibinfo  {journal} {Phys. Rev. Lett.}\ }\textbf {\bibinfo {volume} {102}},\
  \bibinfo {pages} {110403} (\bibinfo {year} {2009})}\BibitemShut {NoStop}%
\bibitem [{\citenamefont {Maceira}\ and\ \citenamefont
  {L{\"a}uchli}(2024)}]{maceira2024thermalization}%
  \BibitemOpen
  \bibfield  {author} {\bibinfo {author} {\bibfnamefont {I.~A.}\ \bibnamefont
  {Maceira}}\ and\ \bibinfo {author} {\bibfnamefont {A.~M.}\ \bibnamefont
  {L{\"a}uchli}},\ }\bibfield  {title} {\bibinfo {title} {Thermalization
  dynamics in closed quantum many body systems: A precision large scale exact
  diagonalization study},\ }\href {https://arxiv.org/abs/2409.18863} {\bibfield
   {journal} {\bibinfo  {journal} {arXiv preprint arXiv:2409.18863}\ }
  (\bibinfo {year} {2024})}\BibitemShut {NoStop}%
\bibitem [{\citenamefont {Matthies}\ \emph {et~al.}(2026)\citenamefont
  {Matthies}, \citenamefont {Dannenfeld}, \citenamefont {Pappalardi},\ and\
  \citenamefont {Rosch}}]{matthies2024thermalization}%
  \BibitemOpen
  \bibfield  {author} {\bibinfo {author} {\bibfnamefont {A.}~\bibnamefont
  {Matthies}}, \bibinfo {author} {\bibfnamefont {N.}~\bibnamefont
  {Dannenfeld}}, \bibinfo {author} {\bibfnamefont {S.}~\bibnamefont
  {Pappalardi}},\ and\ \bibinfo {author} {\bibfnamefont {A.}~\bibnamefont
  {Rosch}},\ }\bibfield  {title} {\bibinfo {title} {Thermalization and
  hydrodynamic long-time tails in a floquet system},\ }\href
  {https://doi.org/10.1103/yz38-zxr9} {\bibfield  {journal} {\bibinfo
  {journal} {Phys. Rev. B}\ }\textbf {\bibinfo {volume} {113}},\ \bibinfo
  {pages} {024305} (\bibinfo {year} {2026})}\BibitemShut {NoStop}%
\bibitem [{non()}]{noncrossing}%
  \BibitemOpen
  \href@noop {} {}\bibinfo {note} {{$\pi$ is a partition (decomposition of
  a set $\{1, \dots q\}$ in blocks) that sums over \emph{non-crossing
  partitions} $NC(n)$, where the blocks do not ``cross'', and
  $\kappa_\pi(t_{n-1}, \dots, t_0) = \kappa_\pi \left ( A(t_{n-1}), \dots,
  A(t_0) \right )$ are products of free cumulants, one for each block of
  $\pi$.}}\BibitemShut {Stop}%
\bibitem [{fir()}]{first-moment}%
  \BibitemOpen
  \href@noop {} {\bibinfo {title} {{ This is a mathematical definition of a
  connected correlation function, valid also for integrable or
  ergodicity-breaking systems. The assumption $\langle A\rangle = 0$ is
  employed only to simplify the expressions, and the main results can be
  straightforwardly generalized to the case $\langle A\rangle \neq 0$
  .}}}\BibitemShut {Stop}%
\bibitem [{\citenamefont {Haehl}\ \emph {et~al.}()\citenamefont {Haehl},
  \citenamefont {Loganayagam}, \citenamefont {Narayan}, \citenamefont
  {Nizami},\ and\ \citenamefont {Rangamani}}]{haehl2017thermal}%
  \BibitemOpen
  \bibfield  {author} {\bibinfo {author} {\bibfnamefont {F.~M.}\ \bibnamefont
  {Haehl}}, \bibinfo {author} {\bibfnamefont {R.}~\bibnamefont {Loganayagam}},
  \bibinfo {author} {\bibfnamefont {P.}~\bibnamefont {Narayan}}, \bibinfo
  {author} {\bibfnamefont {A.~A.}\ \bibnamefont {Nizami}},\ and\ \bibinfo
  {author} {\bibfnamefont {M.}~\bibnamefont {Rangamani}},\ }\bibfield  {title}
  {\bibinfo {title} {Thermal out-of-time-order correlators, KMS relations, and
  spectral functions},\ }\href {https://doi.org/10.1007/JHEP12(2017)154}
  {\bibfield  {journal} {\bibinfo  {journal} {J. High Energy Phys.}}
  {\bibinfo {volume} {12}}\bibinfo  {number} { (2017)}\ \bibinfo {pages}
  {154}}\BibitemShut {NoStop}%
\bibitem [{foo({\natexlab{b}})}]{footnote1}%
  \BibitemOpen
\bibfield  {number} {  }\href@noop {} {} \bibinfo {note}
  {Classical cumulants are the standard connected \ensuremath{n}-point
  function, usually defined by the expansion of the cumulant-generating
  function, but which admit a combinatorial definition in terms of the
  partition set \ensuremath{P(n)}: \ensuremath{\langle \hat A(t_{n-1}) \dots
  \hat A(t_0) \rangle = \sum_{\pi \in P(n)} \langle \hat A(t_{n-1}) \dots \hat
  A(t_0) \rangle_{\pi,c}}, where each \ensuremath{\langle \cdots
  \rangle_{\pi,c}} is a product of classical cumulants corresponding to the
  blocks of the partition \ensuremath{\pi}.}\BibitemShut {Stop}%
\bibitem [{\citenamefont {Leviatan}\ \emph {et~al.}(2017)\citenamefont
  {Leviatan}, \citenamefont {Pollmann}, \citenamefont {Bardarson},
  \citenamefont {Huse},\ and\ \citenamefont {Altman}}]{leviatan2017quantum}%
  \BibitemOpen
  \bibfield  {author} {\bibinfo {author} {\bibfnamefont {E.}~\bibnamefont
  {Leviatan}}, \bibinfo {author} {\bibfnamefont {F.}~\bibnamefont {Pollmann}},
  \bibinfo {author} {\bibfnamefont {J.~H.}\ \bibnamefont {Bardarson}}, \bibinfo
  {author} {\bibfnamefont {D.~A.}\ \bibnamefont {Huse}},\ and\ \bibinfo
  {author} {\bibfnamefont {E.}~\bibnamefont {Altman}},\ }\bibfield  {title}
  {\bibinfo {title} {Quantum thermalization dynamics with matrix-product
  states},\ }\href {https://arxiv.org/abs/1702.08894} {\bibfield  {journal}
  {\bibinfo  {journal} {arXiv preprint arXiv:1702.08894}\ } (\bibinfo {year}
  {2017})}\BibitemShut {NoStop}%
\bibitem [{\citenamefont {Kvorning}\ \emph {et~al.}(2022)\citenamefont
  {Kvorning}, \citenamefont {Herviou},\ and\ \citenamefont
  {Bardarson}}]{klein2022time}%
  \BibitemOpen
  \bibfield  {author} {\bibinfo {author} {\bibfnamefont {T.~K.}\ \bibnamefont
  {Kvorning}}, \bibinfo {author} {\bibfnamefont {L.}~\bibnamefont {Herviou}},\
  and\ \bibinfo {author} {\bibfnamefont {J.~H.}\ \bibnamefont {Bardarson}},\
  }\bibfield  {title} {\bibinfo {title} {{Time-evolution of local information:
  thermalization dynamics of local observables}},\ }\href
  {https://doi.org/10.21468/SciPostPhys.13.4.080} {\bibfield  {journal}
  {\bibinfo  {journal} {SciPost Phys.}\ }\textbf {\bibinfo {volume} {13}},\
  \bibinfo {pages} {080} (\bibinfo {year} {2022})}\BibitemShut {NoStop}%
\bibitem [{\citenamefont {Rakovszky}\ \emph {et~al.}(2022)\citenamefont
  {Rakovszky}, \citenamefont {von Keyserlingk},\ and\ \citenamefont
  {Pollmann}}]{rakowsky2022dissipation}%
  \BibitemOpen
  \bibfield  {author} {\bibinfo {author} {\bibfnamefont {T.}~\bibnamefont
  {Rakovszky}}, \bibinfo {author} {\bibfnamefont {C.~W.}\ \bibnamefont {von
  Keyserlingk}},\ and\ \bibinfo {author} {\bibfnamefont {F.}~\bibnamefont
  {Pollmann}},\ }\bibfield  {title} {\bibinfo {title} {Dissipation-assisted
  operator evolution method for capturing hydrodynamic transport},\ }\href
  {https://doi.org/10.1103/PhysRevB.105.075131} {\bibfield  {journal} {\bibinfo
   {journal} {Phys. Rev. B}\ }\textbf {\bibinfo {volume} {105}},\ \bibinfo
  {pages} {075131} (\bibinfo {year} {2022})}\BibitemShut {NoStop}%
\bibitem [{\citenamefont {Mori}(2024)}]{Mori:2023qbd}%
  \BibitemOpen
  \bibfield  {author} {\bibinfo {author} {\bibfnamefont {T.}~\bibnamefont
  {Mori}},\ }\bibfield  {title} {\bibinfo {title} {Liouvillian-gap analysis of
  open quantum many-body systems in the weak dissipation limit},\ }\href
  {https://doi.org/10.1103/PhysRevB.109.064311} {\bibfield  {journal} {\bibinfo
   {journal} {Phys. Rev. B}\ }\textbf {\bibinfo {volume} {109}},\ \bibinfo
  {pages} {064311} (\bibinfo {year} {2024})}\BibitemShut {NoStop}%
\bibitem [{\citenamefont {Artiaco}\ \emph {et~al.}(2024)\citenamefont
  {Artiaco}, \citenamefont {Fleckenstein}, \citenamefont {Aceituno~Ch\'avez},
  \citenamefont {Kvorning},\ and\ \citenamefont
  {Bardarson}}]{artiaco2024efficient}%
  \BibitemOpen
  \bibfield  {author} {\bibinfo {author} {\bibfnamefont {C.}~\bibnamefont
  {Artiaco}}, \bibinfo {author} {\bibfnamefont {C.}~\bibnamefont
  {Fleckenstein}}, \bibinfo {author} {\bibfnamefont {D.}~\bibnamefont
  {Aceituno~Ch\'avez}}, \bibinfo {author} {\bibfnamefont {T.~K.}\ \bibnamefont
  {Kvorning}},\ and\ \bibinfo {author} {\bibfnamefont {J.~H.}\ \bibnamefont
  {Bardarson}},\ }\bibfield  {title} {\bibinfo {title} {Efficient large-scale
  many-body quantum dynamics via local-information time evolution},\ }\href
  {https://doi.org/10.1103/PRXQuantum.5.020352} {\bibfield  {journal} {\bibinfo
   {journal} {PRX Quantum}\ }\textbf {\bibinfo {volume} {5}},\ \bibinfo {pages}
  {020352} (\bibinfo {year} {2024})}\BibitemShut {NoStop}%
\bibitem [{\citenamefont {Richter}\ \emph
  {et~al.}(2020{\natexlab{b}})\citenamefont {Richter}, \citenamefont
  {Heitmann},\ and\ \citenamefont
  {Steinigeweg}}]{jonas-Ising-10.21468/SciPostPhys.9.3.031}%
  \BibitemOpen
  \bibfield  {author} {\bibinfo {author} {\bibfnamefont {J.}~\bibnamefont
  {Richter}}, \bibinfo {author} {\bibfnamefont {T.}~\bibnamefont {Heitmann}},\
  and\ \bibinfo {author} {\bibfnamefont {R.}~\bibnamefont {Steinigeweg}},\
  }\bibfield  {title} {\bibinfo {title} {{Quantum quench dynamics in the
  transverse-field Ising model: A numerical expansion in linked rectangular
  clusters}},\ }\href {https://doi.org/10.21468/SciPostPhys.9.3.031} {\bibfield
   {journal} {\bibinfo  {journal} {SciPost Phys.}\ }\textbf {\bibinfo {volume}
  {9}},\ \bibinfo {pages} {031} (\bibinfo {year}
  {2020}{\natexlab{b}})}\BibitemShut {NoStop}%
\bibitem [{\citenamefont {Yi-Thomas}\ \emph {et~al.}(2024)\citenamefont
  {Yi-Thomas}, \citenamefont {Ware}, \citenamefont {Sau},\ and\ \citenamefont
  {White}}]{white-Ising-PhysRevB.110.134308}%
  \BibitemOpen
  \bibfield  {author} {\bibinfo {author} {\bibfnamefont {S.}~\bibnamefont
  {Yi-Thomas}}, \bibinfo {author} {\bibfnamefont {B.}~\bibnamefont {Ware}},
  \bibinfo {author} {\bibfnamefont {J.~D.}\ \bibnamefont {Sau}},\ and\ \bibinfo
  {author} {\bibfnamefont {C.~D.}\ \bibnamefont {White}},\ }\bibfield  {title}
  {\bibinfo {title} {Comparing numerical methods for hydrodynamics in a
  one-dimensional lattice spin model},\ }\href
  {https://doi.org/10.1103/PhysRevB.110.134308} {\bibfield  {journal} {\bibinfo
   {journal} {Phys. Rev. B}\ }\textbf {\bibinfo {volume} {110}},\ \bibinfo
  {pages} {134308} (\bibinfo {year} {2024})}\BibitemShut {NoStop}%
\bibitem [{num()}]{numerics}%
  \BibitemOpen
  \href@noop {} {}\bibinfo {note} {Additional results for the Floquet XXZ and
  spin-$1/2$ Ising models are given in the End Matter and Supplemental
  Material.}\BibitemShut {Stop}%
\bibitem [{sin()}]{single-precesion}%
  \BibitemOpen
  \href@noop {} {}\bibinfo {note} {The numerical simulation for $L=19$ is
  performed in single precision.}\BibitemShut {Stop}%
\bibitem [{\citenamefont {Lee}\ \emph {et~al.}(1995)\citenamefont {Lee},
  \citenamefont {Levitov},\ and\ \citenamefont {Yakovets}}]{PhysRevB.51.4079}%
  \BibitemOpen
  \bibfield  {author} {\bibinfo {author} {\bibfnamefont {H.}~\bibnamefont
  {Lee}}, \bibinfo {author} {\bibfnamefont {L.~S.}\ \bibnamefont {Levitov}},\
  and\ \bibinfo {author} {\bibfnamefont {A.~Y.}\ \bibnamefont {Yakovets}},\
  }\bibfield  {title} {\bibinfo {title} {Universal statistics of transport in
  disordered conductors},\ }\href {https://doi.org/10.1103/PhysRevB.51.4079}
  {\bibfield  {journal} {\bibinfo  {journal} {Phys. Rev. B}\ }\textbf {\bibinfo
  {volume} {51}},\ \bibinfo {pages} {4079} (\bibinfo {year}
  {1995})}\BibitemShut {NoStop}%
\bibitem [{\citenamefont {Bertini}\ \emph {et~al.}(2005)\citenamefont
  {Bertini}, \citenamefont {De~Sole}, \citenamefont {Gabrielli}, \citenamefont
  {Jona-Lasinio},\ and\ \citenamefont {Landim}}]{PhysRevLett.94.030601}%
  \BibitemOpen
  \bibfield  {author} {\bibinfo {author} {\bibfnamefont {L.}~\bibnamefont
  {Bertini}}, \bibinfo {author} {\bibfnamefont {A.}~\bibnamefont {De~Sole}},
  \bibinfo {author} {\bibfnamefont {D.}~\bibnamefont {Gabrielli}}, \bibinfo
  {author} {\bibfnamefont {G.}~\bibnamefont {Jona-Lasinio}},\ and\ \bibinfo
  {author} {\bibfnamefont {C.}~\bibnamefont {Landim}},\ }\bibfield  {title}
  {\bibinfo {title} {Current fluctuations in stochastic lattice gases},\ }\href
  {https://doi.org/10.1103/PhysRevLett.94.030601} {\bibfield  {journal}
  {\bibinfo  {journal} {Phys. Rev. Lett.}\ }\textbf {\bibinfo {volume} {94}},\
  \bibinfo {pages} {030601} (\bibinfo {year} {2005})}\BibitemShut {NoStop}%
\bibitem [{fur()}]{further-numerics}%
  \BibitemOpen
  \href@noop {} {}\bibinfo {note} {{To further illustrate our hydrodynamic
  predictions, in the End Matter and Supplemental Material~\cite{SM} we present
  numerical simulations for additional observables, as well as for two other
  one-dimensional spin chains: a Floquet XXZ model and a spin-$1/2$ Ising
  model. Local observables show agreement with the hydrodynamic predictions up
  to finite-size effects, whereas extended observables, e.g., the global
  current, exhibit a fourth-order hydrodynamic tail that is more difficult to
  attribute to finite-size corrections.}}\BibitemShut {Stop}%
\bibitem [{gam()}]{gamma-scaling}%
  \BibitemOpen
  \href@noop {} {}\bibinfo {note} {{ Resolving the exact scaling of
  $\Gamma$ with $L$ would require simulations at much larger system sizes and
  is beyond the scope of this work. In Fig.~\ref{fig:e3} of the End Matter we
  show the second ETH free cumulants in the Floquet XXZ model up to system size
  $L=32$, where a clearer $\Gamma \sim 1/L^2$ scaling is
  observed.}}\BibitemShut {Stop}%
\bibitem [{\citenamefont {Fava}\ \emph {et~al.}(2021)\citenamefont {Fava},
  \citenamefont {Biswas}, \citenamefont {Gopalakrishnan}, \citenamefont
  {Vasseur},\ and\ \citenamefont
  {Parameswaran}}]{doi:10.1073/pnas.2106945118-nl-int}%
  \BibitemOpen
  \bibfield  {author} {\bibinfo {author} {\bibfnamefont {M.}~\bibnamefont
  {Fava}}, \bibinfo {author} {\bibfnamefont {S.}~\bibnamefont {Biswas}},
  \bibinfo {author} {\bibfnamefont {S.}~\bibnamefont {Gopalakrishnan}},
  \bibinfo {author} {\bibfnamefont {R.}~\bibnamefont {Vasseur}},\ and\ \bibinfo
  {author} {\bibfnamefont {S.~A.}\ \bibnamefont {Parameswaran}},\ }\bibfield
  {title} {\bibinfo {title} {Hydrodynamic nonlinear response of interacting
  integrable systems},\ }\href {https://doi.org/10.1073/pnas.2106945118}
  {\bibfield  {journal} {\bibinfo  {journal} {Proc. Natl. Acad. Sci.}\ }\textbf
  {\bibinfo {volume} {118}},\ \bibinfo {pages} {e2106945118} (\bibinfo {year}
  {2021})}\BibitemShut {NoStop}%
\bibitem [{\citenamefont {Wang}(2025)}]{data_wang}%
  \BibitemOpen
  \bibfield  {author} {\bibinfo {author} {\bibfnamefont {J.}~\bibnamefont
  {Wang}},\ }\bibfield  {title} {\bibinfo {title} {Dataset for ``Eigenstate
  thermalization hypothesis correlations via non-linear hydrodynamics"},\
  }\href {https://doi.org/10.5281/zenodo.18010096} {10.5281/zenodo.18010096}
  (\bibinfo {year} {2026})\BibitemShut {NoStop}%
\bibitem [{SM()}]{SM}%
  \BibitemOpen
  \href@noop {} {}\bibinfo {note} {See Supplemental Material for
  additional numerical results for the spin-$1$ Ising model, results for the
  spin-$1/2$ Ising model, as well as details of the numerical method based on
  dynamical quantum typicality, which includes Refs.~[85-86]. }\BibitemShut {Stop}%
\bibitem [{\citenamefont {Ljubotina}\ \emph {et~al.}(2019)\citenamefont
  {Ljubotina}, \citenamefont {Zadnik},\ and\ \citenamefont
  {Prosen}}]{PhysRevLett.122.150605-FXXZ}%
  \BibitemOpen
  \bibfield  {author} {\bibinfo {author} {\bibfnamefont {M.}~\bibnamefont
  {Ljubotina}}, \bibinfo {author} {\bibfnamefont {L.}~\bibnamefont {Zadnik}},\
  and\ \bibinfo {author} {\bibfnamefont {T.}~\bibnamefont {Prosen}},\
  }\bibfield  {title} {\bibinfo {title} {Ballistic spin transport in a
  periodically driven integrable quantum system},\ }\href
  {https://doi.org/10.1103/PhysRevLett.122.150605} {\bibfield  {journal}
  {\bibinfo  {journal} {Phys. Rev. Lett.}\ }\textbf {\bibinfo {volume} {122}},\
  \bibinfo {pages} {150605} (\bibinfo {year} {2019})}\BibitemShut {NoStop}%
\bibitem [{\citenamefont {Lehner}(2002)}]{lehner2002free}%
  \BibitemOpen
  \bibfield  {author} {\bibinfo {author} {\bibfnamefont {F.}~\bibnamefont
  {Lehner}},\ }\bibfield  {title} {\bibinfo {title} {Free cumulants and
  enumeration of connected partitions},\ }\href
  {https://doi.org/10.1006/eujc.2002.0619} {\bibfield  {journal} {\bibinfo
  {journal} {Eur. J. Combin.}\ }\textbf {\bibinfo {volume} {23}},\ \bibinfo
  {pages} {1025} (\bibinfo {year} {2002})}\BibitemShut {NoStop}%
\bibitem [{\citenamefont {Arizmendi}\ \emph {et~al.}(2015)\citenamefont
  {Arizmendi}, \citenamefont {Hasebe}, \citenamefont {Lehner},\ and\
  \citenamefont {Vargas}}]{arizmendi2015relations}%
  \BibitemOpen
  \bibfield  {author} {\bibinfo {author} {\bibfnamefont {O.}~\bibnamefont
  {Arizmendi}}, \bibinfo {author} {\bibfnamefont {T.}~\bibnamefont {Hasebe}},
  \bibinfo {author} {\bibfnamefont {F.}~\bibnamefont {Lehner}},\ and\ \bibinfo
  {author} {\bibfnamefont {C.}~\bibnamefont {Vargas}},\ }\bibfield  {title}
  {\bibinfo {title} {Relations between cumulants in noncommutative
  probability},\ }\href {https://doi.org/10.1016/j.aim.2015.03.029} {\bibfield
  {journal} {\bibinfo  {journal} {Adv. Math.}\ }\textbf {\bibinfo {volume}
  {282}},\ \bibinfo {pages} {56} (\bibinfo {year} {2015})}\BibitemShut
  {NoStop}%
\bibitem [{Note1()}]{Note1}%
  \BibitemOpen
  \bibinfo {note} {For example, $\{\{1,3\},\{2,4\},\{5,6\}\}$ is crossing but
  not connected. On the other hand, $\{\{1,4\},\{2,5\},\{3,6\}\}$ is a
  connected partition.}\BibitemShut {Stop}%
\bibitem [{\citenamefont {Jin}\ \emph {et~al.}(2010)\citenamefont {Jin},
  \citenamefont {De~Raedt}, \citenamefont {Yuan}, \citenamefont
  {I.~Katsnelson}, \citenamefont {Miyashita},\ and\ \citenamefont
  {Michielsen}}]{Jin10-Chebyshev}%
  \BibitemOpen
  \bibfield  {author} {\bibinfo {author} {\bibfnamefont {F.}~\bibnamefont
  {Jin}}, \bibinfo {author} {\bibfnamefont {H.}~\bibnamefont {De~Raedt}},
  \bibinfo {author} {\bibfnamefont {S.}~\bibnamefont {Yuan}}, \bibinfo {author}
  {\bibfnamefont {M.}~\bibnamefont {I.~Katsnelson}}, \bibinfo {author}
  {\bibfnamefont {S.}~\bibnamefont {Miyashita}},\ and\ \bibinfo {author}
  {\bibfnamefont {K.}~\bibnamefont {Michielsen}},\ }\bibfield  {title}
  {\bibinfo {title} {Approach to equilibrium in nano-scale systems at finite
  temperature},\ }\href {https://doi.org/10.1143/JPSJ.79.124005} {\bibfield
  {journal} {\bibinfo  {journal} {J. Phys. Soc. Jpn.}\ }\textbf {\bibinfo
  {volume} {79}},\ \bibinfo {pages} {124005} (\bibinfo {year}
  {2010})}\BibitemShut {NoStop}%
\bibitem [{\citenamefont {De~Raedt}\ and\ \citenamefont
  {Michielsen}(2004)}]{Raedt04}%
  \BibitemOpen
  \bibfield  {author} {\bibinfo {author} {\bibfnamefont {H.}~\bibnamefont
  {De~Raedt}}\ and\ \bibinfo {author} {\bibfnamefont {K.}~\bibnamefont
  {Michielsen}},\ }\bibfield  {title} {\bibinfo {title} {Computational methods
  for simulating quantum computers},\ }\href
  {https://arxiv.org/abs/quant-ph/0406210} {\bibfield  {journal} {\bibinfo
  {journal} {arXiv preprint quant-ph/0406210}\ } (\bibinfo {year}
  {2004})}\BibitemShut {NoStop}%
\end{thebibliography}%

\clearpage

\setcounter{section}{0}
\setcounter{secnumdepth}{2}

\section*{End Matter}
In the End Matter, we present numerical results for a Floquet XXZ model, introduce the relation between classical cumulants and free cumulants, and include a discussion of third-order cumulants in the frequency domain.

\begin{figure}[H]
	\centering
\includegraphics[width=1 \linewidth]{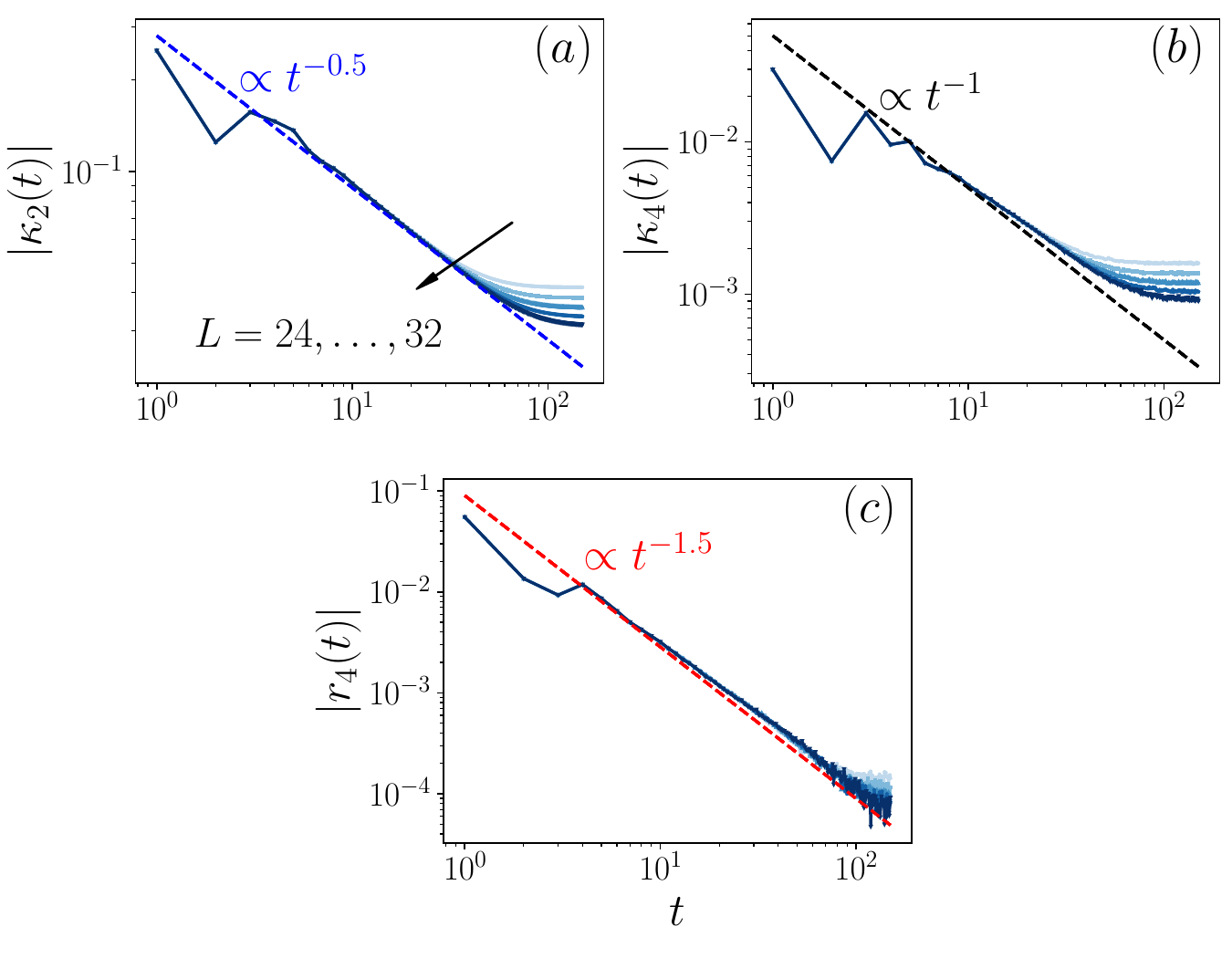}
	\caption{Free cumulants $\kappa_n(t)$ and classical cumulants $r_n(t)$ versus time $t$ for operator ${\hat A} = \hat{\sigma}_\ell^z$ and $\mu = 0$, for $L=24,26,28,30,32$ (from light to dark blue) in the Floquet XXZ model. Blue, black and red dashed line indicate the hydrodynamic prediction $\propto t^{-m/2}$, where $m = 1$ in [(a)], $m=2$ in (b) and $m=3$ in (c).}
	\label{fig:E1}
\end{figure}

\subsection*{Numerical results in the Floquet XXZ model}
To further check our analytical predictions, here we study the cumulants in a Floquet XXZ model with next-nearest coupling. The Floquet operator reads,
$\hat{\cal U}=\hat{\cal U}^{\prime\prime}\hat{\cal U}_{\text{odd}}\hat{\cal U}_{\text{even}}$.
Here
\begin{equation}
{\cal \hat{U}}_{\text{odd}}=\prod_{\ell=1}^{L/2}\hat{U}_{2\ell,2\ell+1}\ ,\ {\cal \hat{U}}_{\text{even}}=\prod_{\ell=1}^{L/2}\hat{U}_{2\ell-1,2\ell},
\end{equation}
where
\begin{equation}
\hat{U}_{\ell,\ell+1}=e^{-i{\cal J}(\hat{\sigma}_{\ell}^{x}\hat{\sigma}_{\ell+1}^{x}+\hat{\sigma}_{\ell}^{y}\hat{\sigma}_{\ell+1}^{y})-i{\cal J}^{\prime}(\hat{\sigma}_{\ell}^{z}\hat{\sigma}_{\ell+1}^{z}-\mathbb{1})},
\end{equation}
and the next-nearest term ${\cal \hat{U}}^{\prime\prime}=e^{-i{\cal J}^{\prime\prime}\sum_{\ell=1}^{L}\hat{\sigma}_{\ell}^{z}\hat{\sigma}_{\ell+2}^{z}}.$
Here, $\sigma^{x,y,z}_\ell$ denote the Pauli matrices acting on site $\ell$.
The total magnetization along the $z$ direction
${\cal S}=\sum_{\ell=1}^{L}\hat{\sigma}_{\ell}^{z}$
is the only conserved quantity.
The eigenstates and quasi-eigenvalues of the Floquet operator are denoted by $|\gamma, k\rangle$ and $\phi_{\gamma k} \in [0, 2\pi]$, respectively,  satisfying 
\begin{equation}
\hat{{\cal U}}|\gamma,k\rangle=e^{-i\phi_{\gamma k}}|\gamma,k\rangle,\ \ \ {\cal \hat{S}}|\gamma,k\rangle=S_{k}|\gamma,k\rangle.
\end{equation}
For brevity, in the following we denote $|\alpha\rangle \equiv |\gamma, k\rangle$, $\phi_\alpha \equiv \phi_{\gamma k}$, and $S_\alpha \equiv S_k$.

\begin{figure}[H]
	\centering
\includegraphics[width=1 \linewidth]{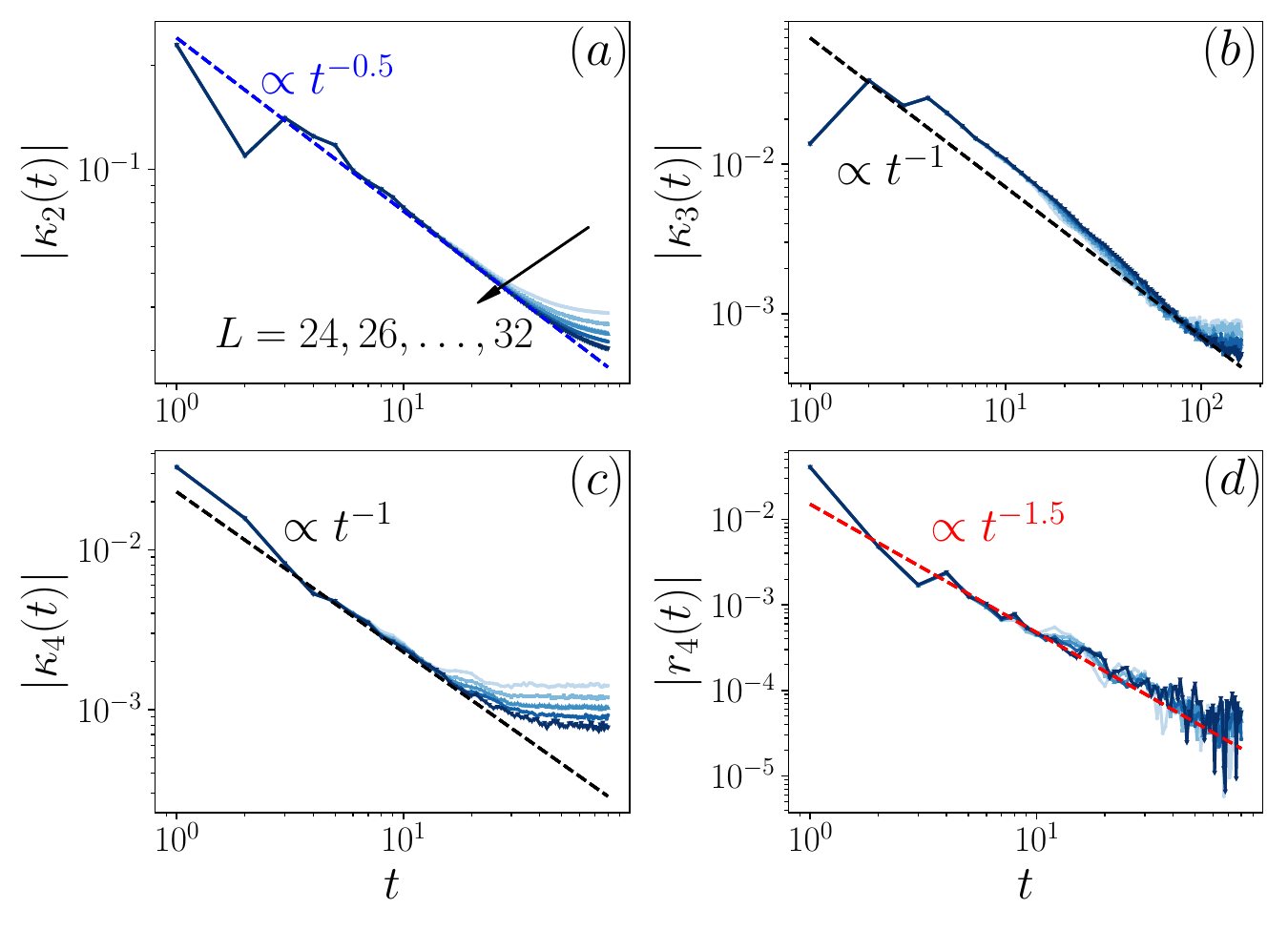}
	\caption{Similar to Fig.~\ref{fig:E1} but for $\mu = 0.3$. Blue, black and red dashed line indicate the hydrodynamic prediction $\propto t^{-m/2}$, where $m = 1$ in [(a)], $m=2$ in [(b)-(c)] and $m=3$ in (d).}
	\label{fig:E2}
\end{figure}

\begin{figure}[H]
\includegraphics[width=1. \linewidth]{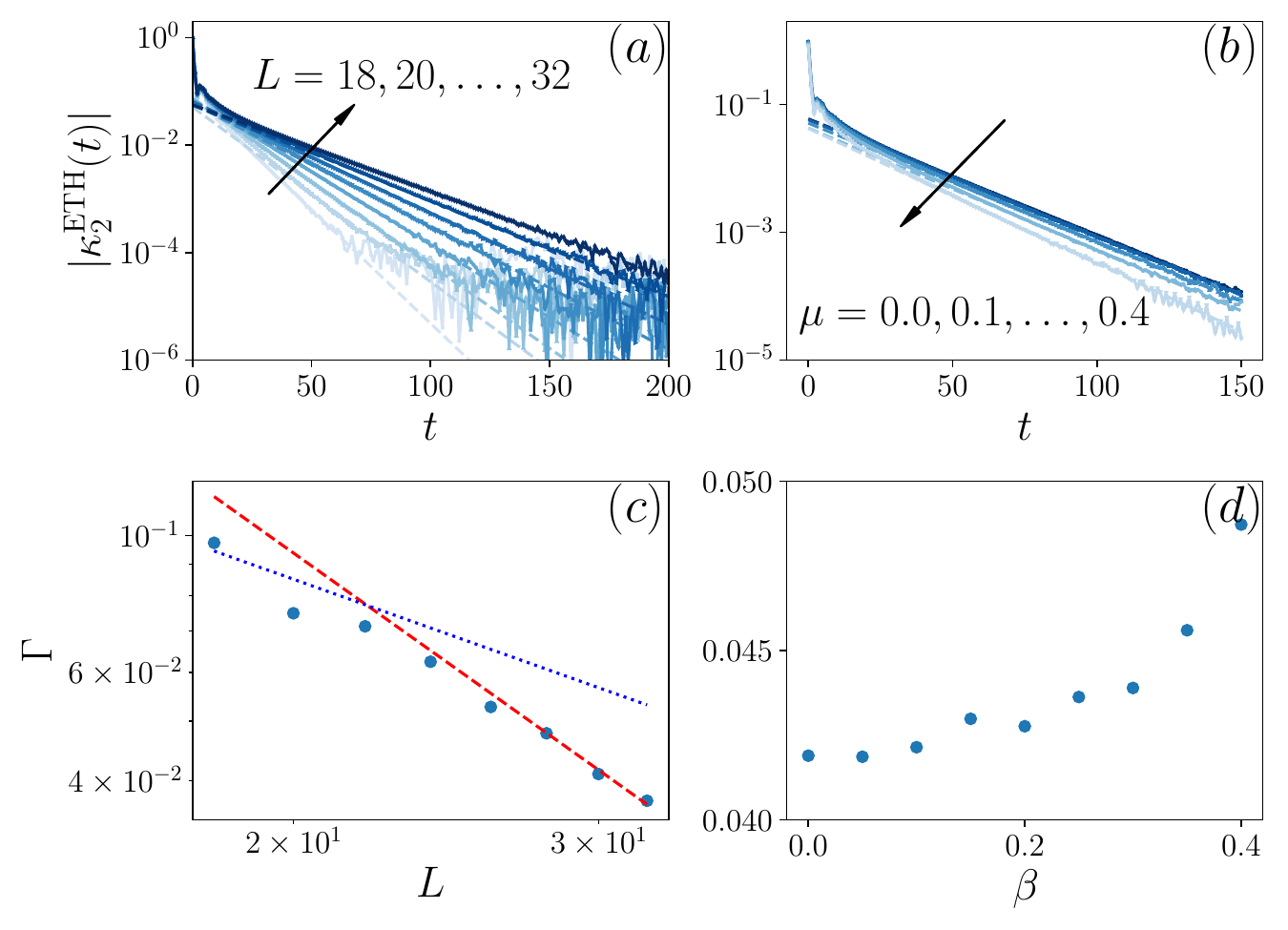}	\caption{ETH free cumulants $\kappa^\text{ETH}_2(t)$ as a function of time $t$ for the operator ${\hat A} = \hat{\sigma}^z_\ell$ for (a) $\mu = 0$ with system sizes $L = 18, 20, \ldots, 32$ (from light to dark blue), and (b) $L = 30$ for $\mu = 0, 0.1, 0.2, 0.3, 0.4$ (from dark to light blue). The dashed line represents the the exponential fit $\propto \exp(-\Gamma t)$. The fitting is performed in the region where $\kappa_{2}^{\text{ETH}}(t)\in[10^{-2.5},10^{-4}]$ and the corresponding fitting parameters are shown in the (c) and (d). The dashed and dotted line in (c) represents the scaling $\propto 1/L^{2}$ and $\propto 1/L$, respectively.}
	\label{fig:e3}
\end{figure}

 When ${\cal J}^{\prime\prime} = 0$ (i.e., in the absence of next-nearest-neighbor coupling), the system is integrable and exhibits various types of spin transport---ballistic, diffusive, or anomalous---depending on the values of ${\cal J}$ and ${\cal J}^\prime$~\cite{PhysRevLett.122.150605-FXXZ}. 
For ${\cal J}^{\prime\prime} \neq 0$, the system becomes non-integrable, and the transport behavior is generally expected to be diffusive. Here, we choose the parameters ${\cal J} = \frac{\pi}{8}$, ${\cal J}^\prime = \frac{\pi}{4}$, and ${\cal J}^{\prime\prime} = \frac{\pi}{8}$ and employ periodic boundary condition.
The free and classical cumulants are calculated from correlation functions $\langle A(t_{n-1})A(t_{n-2})\ldots A\rangle$ using Eqs.~\eqref{eq_freecumbeta} and ~\eqref{eq:kappa-to-c}, where  $\langle\cdot\rangle=\text{tr}(e^{-\mu{\cal S}}\cdot)/Z$ with $Z=\text{tr}(e^{-\mu{\cal S}})$. The ETH free cumulants are defined as
\begin{equation}
  \kappa_{n}^{\text{ETH}}(\vec{t})=\frac{1}{Z}\sum_{\alpha_{1}\neq\ldots\neq\alpha_{n}}e^{-\mu S_{\alpha_{1}}}e^{i\vec{\omega}\cdot\vec{t}}A_{\alpha_{1}\alpha_{2}}\ldots A_{\alpha_{n}\alpha_{1}},
\end{equation}
where $\vec t = (t_{n-1}, \dots t_1)$, $\vec \omega=(\omega_{n-1}, \dots, \omega_1) $ with $\omega_{n}=\phi_{\alpha_{n-1}}-\phi_{\alpha_{n-2}}$.

The free and classical cumulants are presented in Fig.~\ref{fig:E1} for $\mu = 0$ and in Fig.~\ref{fig:E2} for $\mu = 0.3$, where we consider system sizes up to $L = 32$. 
With increasing system size, the results show a clear tendency to converge toward the predicted scaling behaviors $\kappa_2(t) \sim t^{-1/2}$, $\kappa_3(t) \sim t^{-1}$, $\kappa_4(t) \sim t^{-1}$, and $r_4(t) \sim t^{-1.5}$.  This further supports the generality of our hydrodynamic prediction given in Eq.~\eqref{eq_cumu_prediction}.

In addition, we also study the ETH free cumulants. Due to particle–hole symmetry, the odd cumulants vanish at $\mu = 0$, only the second ETH free cumulant is studied here.
The results  are shown in Fig.~\ref{fig:e3},  where a late-time exponential decay $\sim e^{-\Gamma t}$ is observed in all cases considered. The scaling $\Gamma \sim 1/L^2$ can be seen for $L>22$, and $\Gamma$ is found to increase roughly with increasing {$\mu$}, consistent with the results in Fig.~\ref{fig:3}.

\subsection*{Relation between classical and free cumulants}
The classical and free cumulants are related by \cite{lehner2002free, arizmendi2015relations}
\begin{equation}
\kappa_n(t_{n-1},\dots,t_0) = \sum_{\pi \in {P_{\rm conn}}(n)} \langle \hat A(t_{n-1})\dots \hat A(t_0)\rangle_{c, \pi}\ , \label{eq:kappa-to-c}
\end{equation}
where $\langle \hat A(t_{n-1})\dots \hat A(t_0)\rangle_{c, \pi}$ represents products of classical cumulants, one for each block of $\pi$ and $P_{\rm conn}(n)$ denotes the set of ``connected partitions'', defined by the condition that no proper subinterval of $\pi$ is
a union of blocks (i.e. the diagram of $\pi$ is a connected graph) \footnote{For example, $\{\{1,3\},\{2,4\},\{5,6\}\}$ is crossing but not connected. On the other hand, $\{\{1,4\},\{2,5\},\{3,6\}\}$ is a connected partition.}.
At the lowest orders this relation reads:
\begin{subequations}\label{eq-cc-fc}
    \begin{align}
        \kappa_{2}(t_1, 0) & = \langle \hat A(t_1) \hat A\rangle_c\\
        \kappa_3(t_2, t_1, 0) & = \langle \hat A(t_2) \hat A(t_1) \hat A\rangle_c \\
\kappa_4(t_3, t_2, t_1, 0) & = \langle \hat A(t_3) \hat A(t_2) \hat A(t_1) \hat A\rangle_c  \nonumber \\
& \quad + \braket{\hat A(t_3) \hat A(t_1) }_c \braket{\hat A(t_2) \hat A }_c\ .
    \end{align}
\end{subequations}

\subsection*{Classical cumulants in frequency domain}
We can use our knowledge of the three-point functions in frequency space to get an expression for the functions $F_{e_\beta}^{(n)}(\vec \omega)$ in \eqref{eq_Kn_eth}. In \cite{delacretaz2024nonlinear}, the leading order three-point function for conserved densities was found: 
\begin{equation}
\label{eq_3point_momspace}
\begin{split}
&\langle h(\omega_3,p_3)h(\omega_2,p_2)h(\omega_1,p_1)\rangle_\beta \\
&=(T\chi)^2 \left[\frac{2 \sigma' }{\sigma}- \frac{4D' }{D}\right] \frac{k_1^2k_2^2k_3^2 (k_1^2 + k_2^2+k_3^2)}
	{(\omega_1^2 + k_1^4)(\omega_2^2 + k_2^4)(\omega_3^2 + k_3^4)}\\ 
	&-(T\chi)^2\frac{4 \sigma'}{\sigma}
	\frac{ k_1^2 k_{23} \omega_2 \omega_3 + 
	k_2^2 k_{31} \omega_3 \omega_1 + 
	k_3^2 k_{12} \omega_1 \omega_2
	}{(\omega_1^2 + k_1^4)(\omega_2^2 + k_2^4)(\omega_3^2 + k_3^4)}\, , 
\end{split}
\end{equation}
where momenta were rescaled as $k_i=\sqrt{D}p_i$, and $p_3=-p_1-p_2$, $\omega_3=-\omega_2-\omega_1$, and we used the shorthand notation $k_{ij}=k_i\cdot k_j$. There are two pieces in the three-point function above -- one proportional to $\sigma'\equiv d\sigma/d \langle h\rangle$ and the other to $D'\equiv d D/d \langle h\rangle$. To obtain the autocorrelation function of local energy densities $h(x)$ at $x=0$, one integrates over the momenta,
\begin{equation}
\label{eq_3pt_FT}
\begin{split}
&\langle h(\omega_3)h(\omega_2)h(\omega_1)\rangle_\beta \\
&\quad = \int \frac{dk_1 dk_2}{(2\pi)^2}
\langle h(\omega_3,k_3)h(\omega_2,k_2)h(\omega_1,k_1)\rangle_\beta
\end{split}
\end{equation}

Performing this integral for the case when $\omega_1>\omega_2>0$, one finds the contribution of both terms as:
\begin{gather}\label{eq_Gs_GD}
\langle h(\omega_3)h(\omega_2)h(\omega_1)\rangle_\beta 
    =  \frac{\chi^2 T^2}{4}\left[\frac{\sigma'}{\sigma}G_{\sigma'} + \frac{D'}{D}G_{D'}\right]\, , \\
G_{\sigma'}=\frac{1}{\sqrt{\omega_1\omega_2}}\, ,
    G_{D'}+G_{\sigma'}=-\frac{2}{\sqrt{\omega_2(\omega_1+\omega_2)}}\, .
\end{gather}
Using this with \eqref{eq-cc-fc} for $\hat{A}=\hat{h}_i$, produces the classical cumulant in frequency space. Assuming the validity of ETH, this then gives us an expression for the function $F_{e_\beta}^{(3)}(\vec{\omega})$ in \eqref{eq_Kn_eth}
\begin{gather}
\label{eq_3pt_freq}
F_{e_\beta}^{(3)}(\vec \omega) e^{-\beta \vec \omega \cdot \vec \ell_3}
= \chi^2 T^2 \pi^2 \cdot \nonumber \\
 \Bigg[
\frac{\sigma'}{\sigma}\frac{1}{\sqrt{\omega_1 \omega_2}} 
\quad - \frac{D'}{D} \left(
\frac{1}{\sqrt{\omega_1 \omega_2}}
+ \frac{2}{\sqrt{\omega_2 (\omega_1 + \omega_2)}}
\right)
\Bigg] \, .
\end{gather}
Note that while the 3-point function can entirely be expressed in terms of time-ordered correlators using KMS relations, this is no longer the case for 4 and higher-point functions. In those cases, the ETH functions $F^{(n)}$ contain information about out-of-time-ordered correlators as well.

\setcounter{figure}{0}
\setcounter{equation}{0}
\renewcommand{\thefigure}{S\arabic{figure}}
\renewcommand{\theequation}{S\arabic{equation}}

\section*{Supplemental Material}

In the Supplemental Material, we present additional numerical results of:
\begin{enumerate}
    \item Free/classical cumulants for different observables and temperatures in spin-$1$ Ising model;
    \item Second ETH free cumulants for energy density observable in the spin-$1$ Ising model;
    \item Free/classical cumulants for energy density observables in the spin-$1/2$ Ising model;
    \item Details on the numerical methods based on dynamical quantum typicality.
\end{enumerate}

\subsection*{Hydrodynamic free-cumulants for different observables in spin-$1$ Ising model}
\begin{figure}[H]
	\centering
	\includegraphics[width=1 \linewidth]{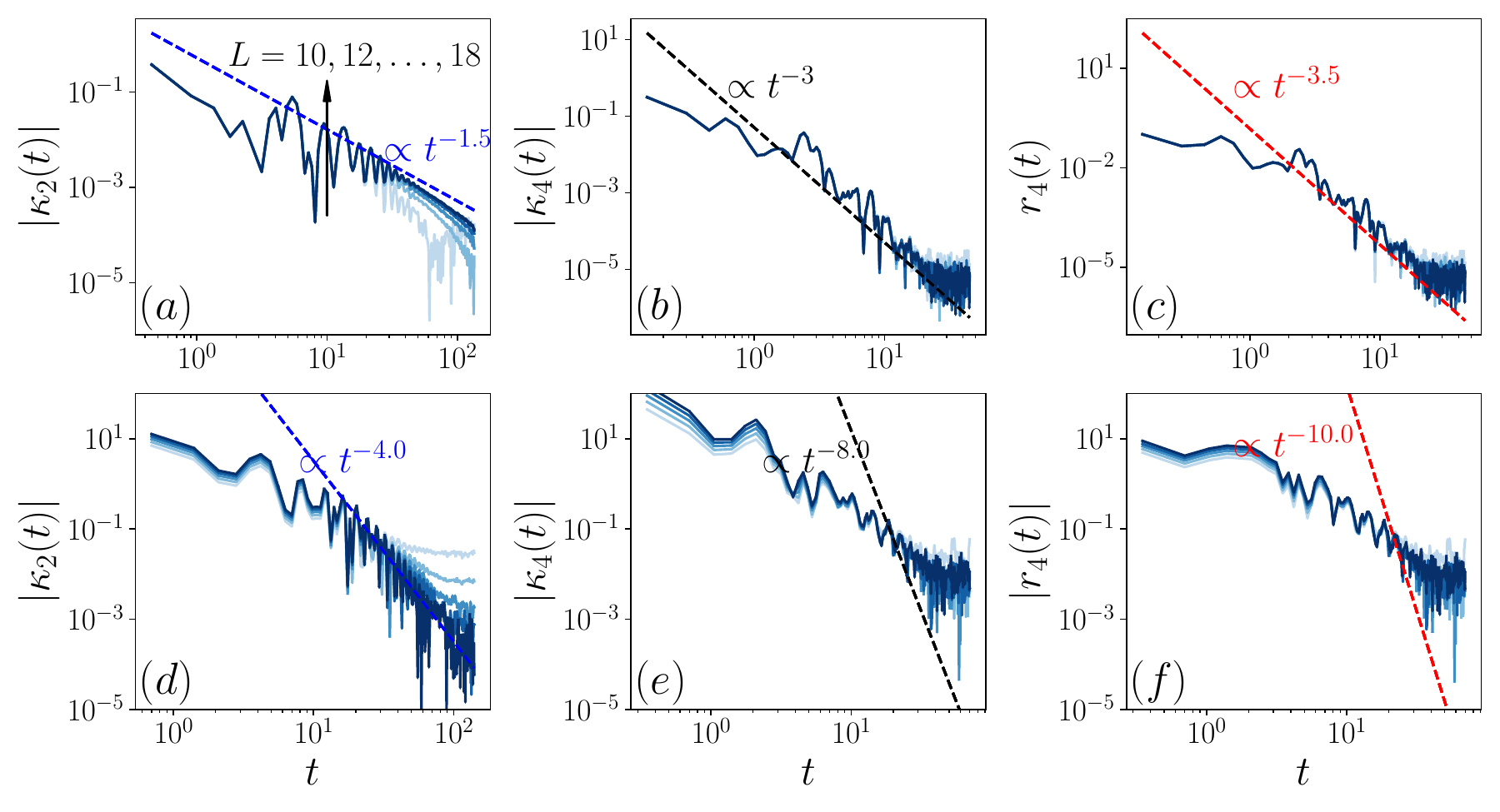}
	\caption{Local and total current:  Free cumulants $\kappa_n(t)$ and classical cumulants $r_n(t)$ versus time $t$ for 
    local current $\hat{j}_i$ [(a)(b)(c)]  and total current $\hat{j}_{\text{tot}}$\  [(d)(e)(f)] for $L=10,12,14,16,18$ (from light to dark blue). The dashed line indicates the hydrodynamic prediction.}
	\label{fig:S1}
\end{figure}

\begin{figure}[H]
	\centering
	\includegraphics[width=1 \linewidth]{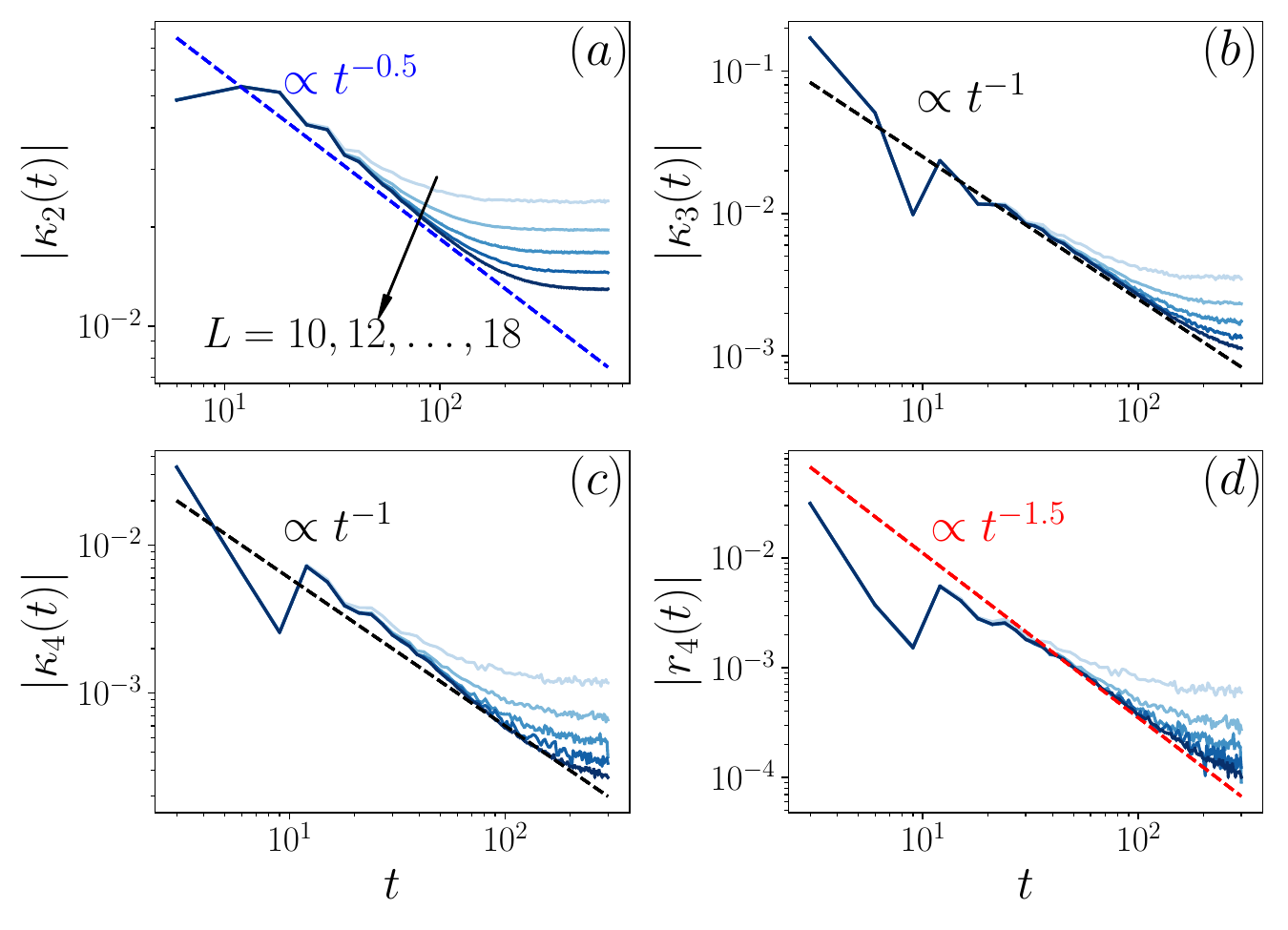}
	\caption{Free cumulants $\kappa_n(t)$ and classical cumulants $r_n(t)$ versus time $t$ for operator ${\hat A} = \hat{s}_i^z$ and $\beta = 0$, for $L=10,12,14,16,18$ (from light to dark blue). Blue, black and red dashed line indicate the hydrodynamic prediction $\propto t^{-m/2}$, where $m = 1$ in [(a)], $m=2$ in [(b)-(c)] and $m=3$ in (d).}
	\label{fig:SZ}
\end{figure}

\begin{figure}[H]
	\centering
	\includegraphics[width=1 \linewidth]{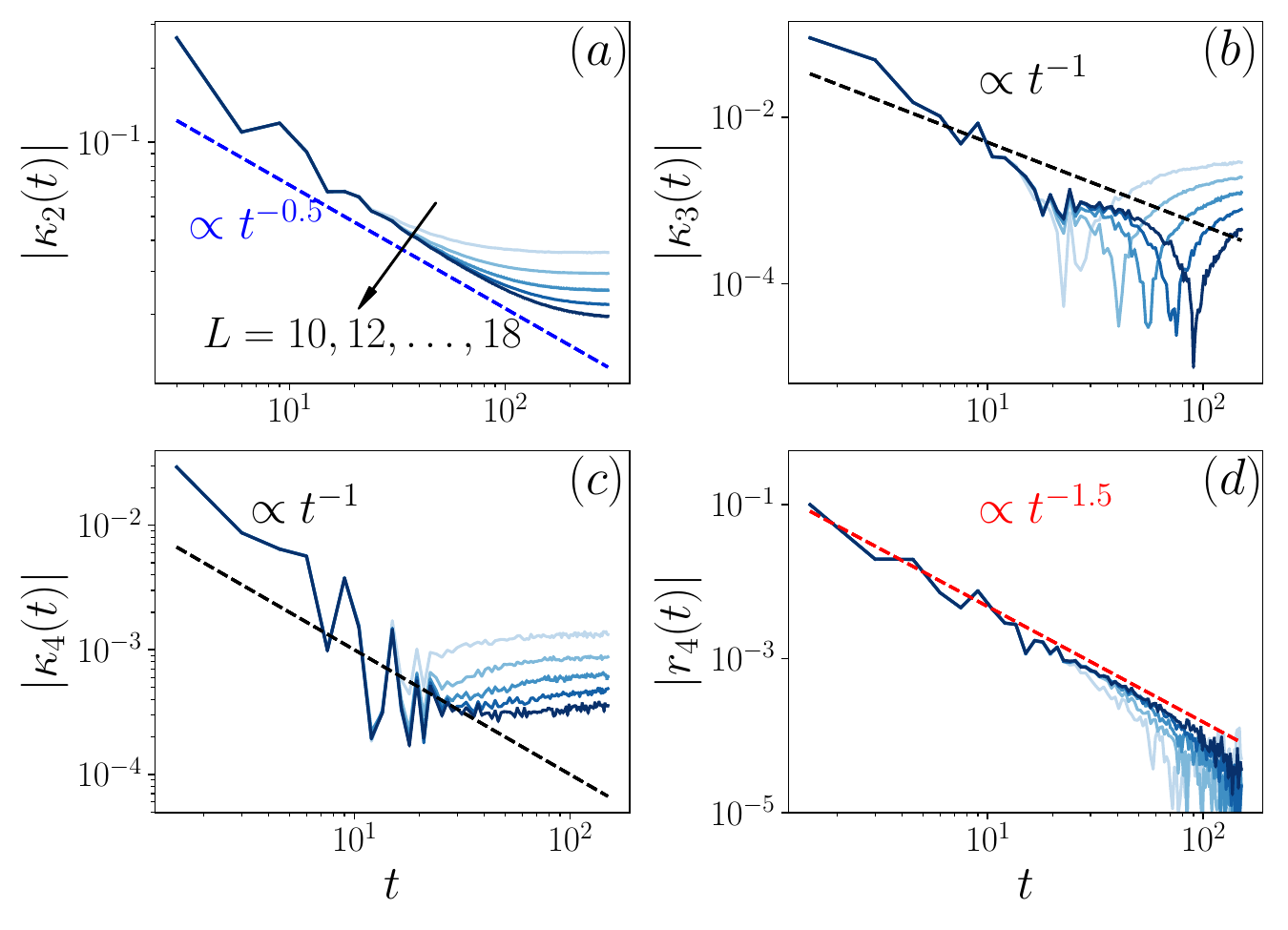}
	\caption{Similar to Fig.~\ref{fig:SZ}, but for operator $\hat{A} = \hat{s}^x_i$. }
	\label{fig:SX}
\end{figure}

To evaluate the generality of our analytical prediction, we present results for both free and classical cumulants of various observables, including local current $\hat{j}_i$ and global current $\hat{j}_\text{tot}=\sum_i \hat{j}_i$ (Fig.~\ref{fig:S1});
$\hat{s}^z_i$ (Fig.~\ref{fig:SZ}) and $\hat{s}^x_i$ (Fig.~\ref{fig:SX}).
\new{Good} agreement with our hydrodynamic prediction is observed for most of the observables considered.   Deviations are visible in $\kappa_4$ and $r_4$ for the total current $\hat{j}_\text{tot}$ [Fig.~\ref{fig:S1}(e,f)] and in $\kappa_4$ for the local spin component $\hat{s}^x_i$ (Fig.~\ref{fig:SX} (c)).
For $\hat{s}^x_i$, the deviation in $\kappa_4$ is likely due to finite-size effects, as the figure shows that convergence has not been reached within the system sizes considered (in comparison to Fig.~\ref{fig:SZ} (c)).
For $\hat{j}_\text{tot}$, the finite-size effects are even more pronounced, as a consequence of its collective nature and rapid decay, which make it significantly more difficult to observe the hydrodynamic prediction—particularly for higher cumulants.  Nevertheless, it is notable that the predicted $t^{-4.0}$ scaling is still clearly visible in the two-point correlation function.

\new{Moreover, in addition to Fig.~2 (b) in the main text, we present results for the operator $\hat{A} = \hat{h}_i$ at finite temperature $\beta = 0.2$ for various system sizes, which suggest a convergence toward the hydrodynamic prediction as the system size increases.}

\begin{figure}[H]
	\centering
	\includegraphics[width=1 \linewidth]{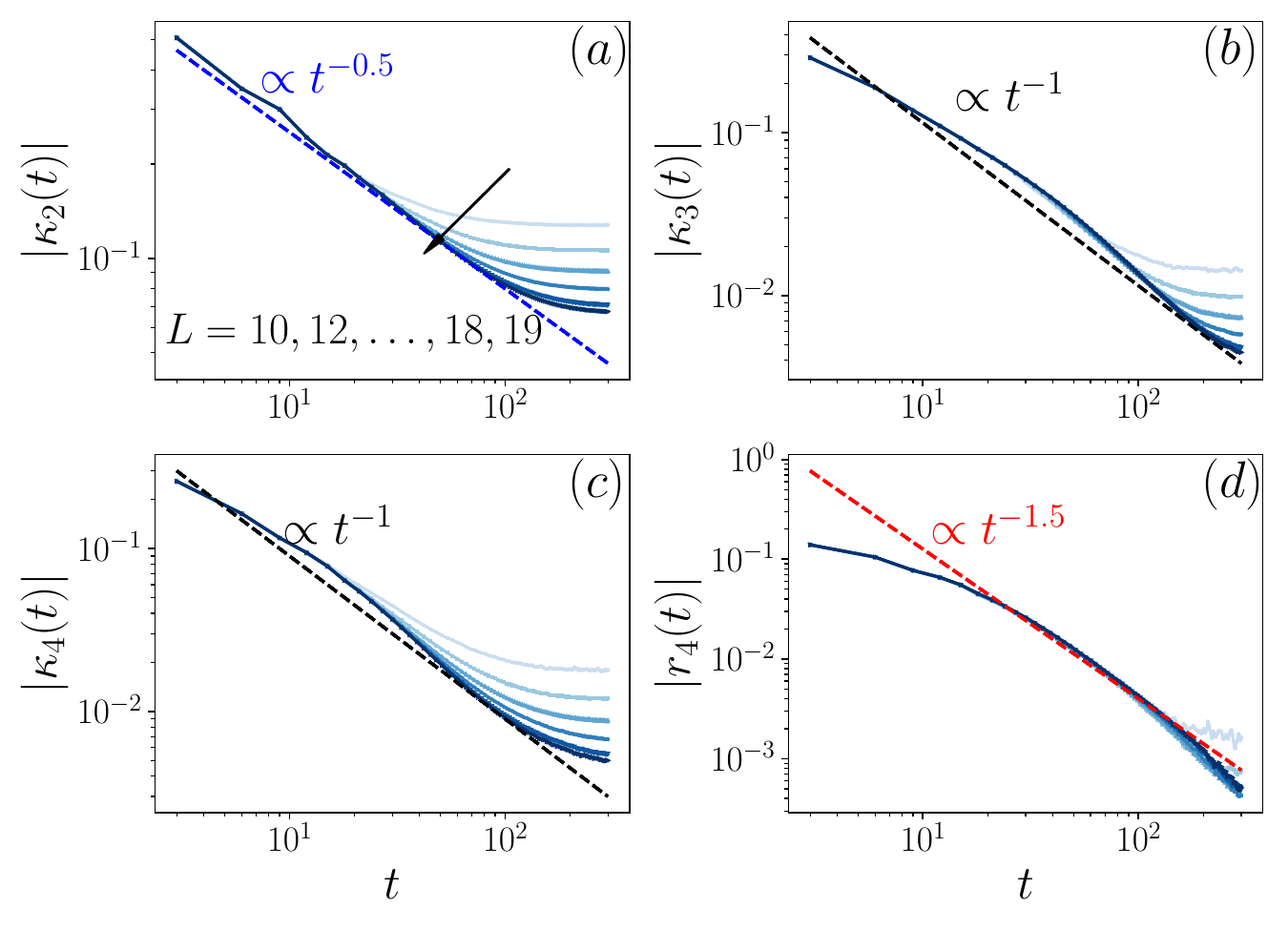}
	\caption{\new{Free cumulants $\kappa_n(t)$ and classical cumulants $r_n(t)$ versus time $t$ for operator ${\hat A} = \hat{h}_i$ and $\beta = 0.2$, for $L=10,12,14,16,18, 19$ (from light to dark blue). Blue, black and red dashed line indicate the hydrodynamic prediction $\propto t^{-m/2}$, where $m = 1$ in [(a)], $m=2$ in [(b)-(c)] and $m=3$ in (d).}}
	\label{fig:s5}
\end{figure}

\subsection*{Finite-size exponential decay for the second ETH free cumulants}

\begin{figure}[H]
	\centering
	\includegraphics[width=1. \linewidth]{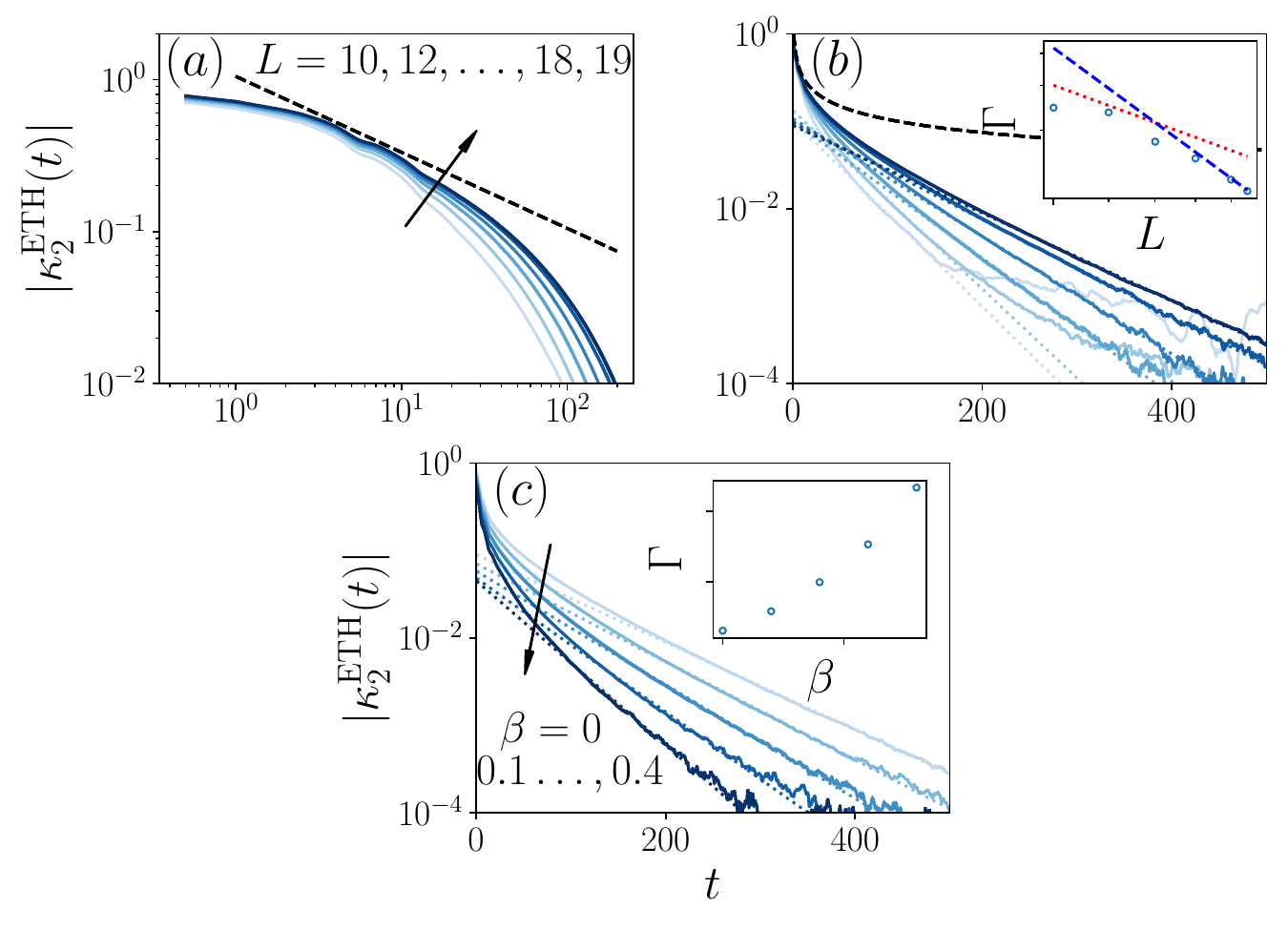}
	\caption{ETH free cumulants $\kappa^\text{ETH}_2(t)$ as a function of time $t$ for the operator ${\hat A} = \hat{h}_i$ for (a, b) $\beta = 0$ with system sizes $L = 10, 12, 14,16, 18,19$ (from light to dark blue), and (c) $L = 19$ for $\beta = 0, 0.1, 0.2, 0.3, 0.4$ (from light to dark blue). 
    The dashed line represents the power-law scaling $\propto t^{-0.5}$, and the dotted line corresponds to the exponential fit $\propto \exp(-\Gamma t)$.
\new{The exponential fitting is performed in the region where $\kappa_{2}^{\mathrm{ETH}}(t) \in [10^{-2.5}, 10^{-4}]$ for $L > 12$, and in $\kappa_{2}^{\mathrm{ETH}}(t) \in [10^{-1.5}, 10^{-2.5}]$ for $L \le 12$.
The corresponding fitting parameters $\Gamma$ are shown in the insets of panel (b) on a double-logarithmic scale and in panel (c) on a linear scale.
}}
	\label{figS:KETH2}
\end{figure}

Analogous to Fig.~3 in the main text, Fig.~\ref{figS:KETH2} presents results for the second ETH cumulant, $\kappa^{\text{ETH}}_2$, of the energy density operator $\hat{A} = \hat{h}_i$. Similar behavior is observed:
1) $\kappa^{\text{ETH}}_2$ exhibits a power-law decay at intermediate times, followed by exponential behavior $\sim e^{-\Gamma t}$ at later times;
2) the decay rate $\Gamma$ decreases with increasing system size and temperature.

\subsection*{Hydrodynamic free cumulants in spin-$\frac{1}{2}$ Ising model}
\begin{figure}[t]
	\centering
	\includegraphics[width=1.0 \linewidth]{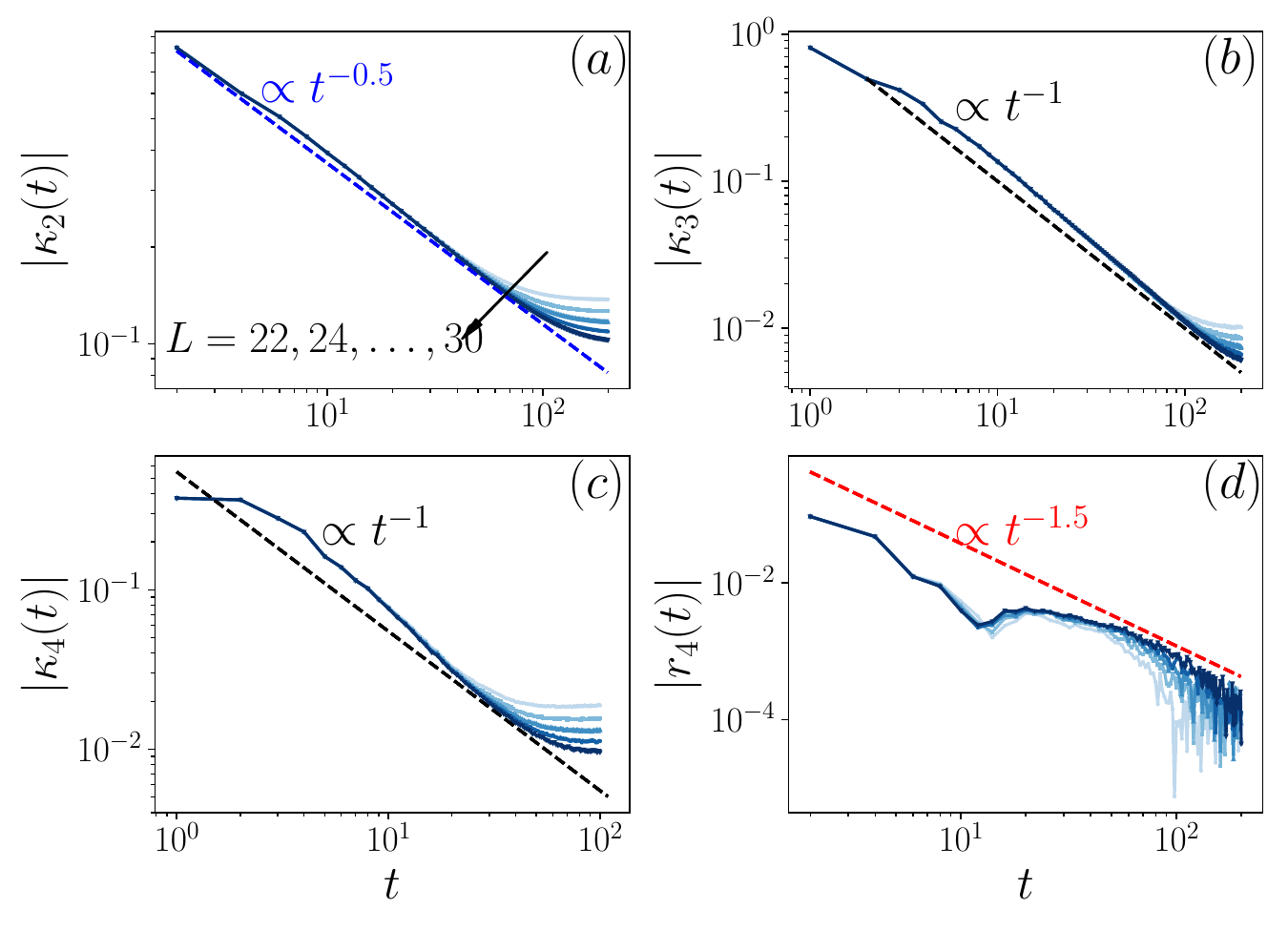}
	\caption{\new{Free cumulants $\kappa_n(t)$ and classical cumulants $r_n(t)$ of the energy density $\hat A = \hat h_i$ versus time $t$ in the spin-$\frac{1}{2}$ Ising model: (a) $\kappa_2(t)$; (b) $\kappa_4(t)$; (c) $\kappa_3 (t)$ and (d) $r_4(t) = \kappa_4(t) - (\kappa_2(2t))^2$ for $L = 22,24,26,28,30$(from light to dark blue), at infinite temperature $\beta = 0$. Blue, black and red dashed lines indicate the hydrodynamic predictions from Eq.~(15) in the main text. }}
	\label{fig:half}
\end{figure}

\begin{figure}[t]
	\centering
	\includegraphics[width=1.0 \linewidth]{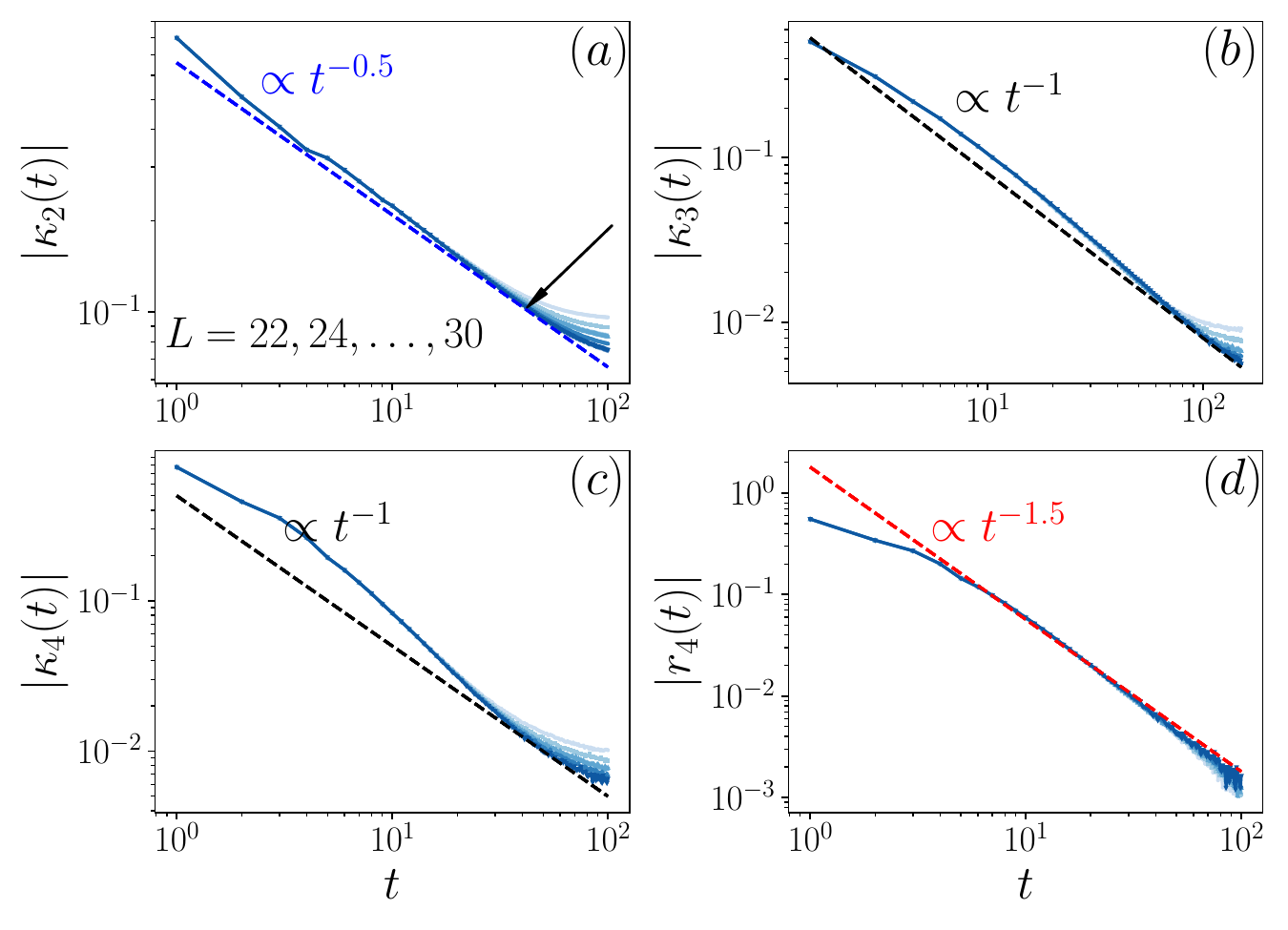}
	\caption{\new{Similar to Fig.~\ref{fig:half}, but at finite temperature $\beta = 0.2$}}
	\label{fig:s7}
\end{figure}

\new{As a further evaluation, we checked our analytical prediction in a spin-$\frac{1}{2}$ mixed field Ising model, with the Hamiltonian $\hat{H}=\sum_{i}\hat{h}_{i}$, where 
\begin{equation}
\hat{h}_{i}=J\hat{\sigma}_{i}^{z}\hat{\sigma}_{i+1}^{z}+\frac{1}{2}g_{z}(\hat{\sigma}_{i}^{z}+\hat{\sigma}_{i+1}^{z})+\frac{1}{2}g_{x}(\hat{\sigma}_{i}^{x}+\hat{\sigma}_{i+1}^{x})
\end{equation}
denotes the energy density. Here $\hat \sigma_i^\mu$ are the Pauli operators in the $\mu=x, z$ direction and we use periodic boundary conditions $\hat{\sigma}_{L+1}^{\mu}=\hat{\sigma}_{1}^{\mu}$. The parameters are fixed as $g_x=1.1$, $g_z=0.9$ and $J = 1.0$. The free and classical cumulants for the local energy density operator $A = \hat{h}_i$ are shown in Fig.~\ref{fig:half} at infinite temperature $\beta = 0$ and in Fig.~\ref{fig:s7} at finite temperature $\beta = 0.2$. A good agreement with the hydrodynamic predictions (given in Eq.~(15) in the main text) is observed, with a tendency toward convergence as the system size increases, further supporting the generality of our results.}

\subsection*{Details on Dynamical Quantum Typicality}

We briefly introduce the numerical method used in the main text to calculate free, classical and ETH free cumulants, which is based on the dynamical quantum typicality (DQT) \cite{DQT-Gemmer}, and check its accuracy by comparing the results to exact diagonalization (ED) results.

\subsubsection{Free cumulants and classical cumulants}
According to Eqs. (3) and (7) in  the main text, the free cumulants $\kappa_n$ and classical cumulants $r_n$ are unambiguously determined by $n$-point correlation function $\langle\hat{A}(t_{n-1})\hat{A}(t_{n-2})\cdots\hat{A}(t_{1})\hat{A}\rangle$.  In the following, we introduce the numerical method for computing this $n$-point correlation function.

Let us start from a normalized Haar random state in the Hilbert space
\begin{equation}
    |\psi\rangle = \sum_j C_j |E_j\rangle,
\end{equation}
where $C_j$ are complex numbers whose real and imaginary parts are drawn independently from a Gaussian distribution. 
According to DQT, the $2$-point function at inverse temperature $\beta$ 
\begin{equation}
    \langle\hat{A}(t)\hat{A}\rangle=\frac{1}{Z}\text{Tr}[e^{-\beta H}\hat{A}(t)\hat{A}]\ ,
\end{equation}
where $Z=\text{Tr}[e^{-\beta H}]$, can be written as
\begin{equation}
\langle\hat{A}(t)\hat{A}\rangle=\frac{1}{\langle\psi|e^{-\beta H}|\psi\rangle}\langle\psi|e^{-\frac{\beta}{2}H}e^{iHt}\hat{A}e^{-iHt}\hat{A}e^{-\frac{\beta}{2}H}|\psi\rangle+\varepsilon(t).
\end{equation}
The error scales as
\begin{equation}
    \varepsilon(t)\sim\frac{1}{\sqrt{D}},
\end{equation}
with $D$ being the Hilbert space dimension of the system.
If $D$ is sufficiently large, one has
\begin{equation}\label{eq-Ct-typ}
\langle\hat{A}(t)\hat{A}\rangle\simeq\frac{1}{\langle\psi|e^{-\beta H}|\psi\rangle}\langle\psi|e^{-\frac{\beta}{2}H}e^{iHt}\hat{A}e^{-iHt}\hat{A}e^{-\frac{\beta}{2}H}|\psi\rangle.
\end{equation}
The accuracy of Eq.~\eqref{eq-Ct-typ} can be further improved by taking an average over $N_p$ different realizations of Haar random state $|\psi\rangle$, and 
\begin{equation}
    \epsilon(t)\sim\frac{1}{\sqrt{D N_p}}.
\end{equation}    
In our simulations, we choose $N_p = M \cdot 3^{L_\text{max}-L}$ ($M \ge 1$), which ensures similar accuracy for different system size $L$. 
Employing two auxiliary states
\begin{equation}
|\psi_{\beta}\rangle=e^{-\frac{\beta}{2}H}|\psi\rangle,\quad|\psi_{\beta}^{A}\rangle=\hat{A}e^{-\frac{\beta}{2}H}|\psi\rangle,
\end{equation}
Eq. \eqref{eq-Ct-typ} can be written as
\begin{equation}
\langle\hat{A}(t)\hat{A}\rangle\simeq\frac{1}{\langle\psi_{\beta}|\psi_{\beta}\rangle}\langle\psi_{\beta}(t)|\hat{A}|\psi_{\beta}^{A}(t)\rangle,
\end{equation}
where
\begin{equation}
|\psi_{\beta}(t)\rangle=e^{-iHt}|\psi_{\beta}\rangle,\quad|\psi_{\beta}^{A}(t)\rangle=e^{-iHt}|\psi_{\beta}^{A}\rangle.
\end{equation}

Similarly, the $n$-point correlation function can be calculated by DQT, using  
\begin{gather}
   \langle\hat{A}(t_{n-1})\hat{A}(t_{n-2})\cdots\hat{A}(t_{1})\hat{A}\rangle \nonumber \\
   \simeq \frac{1}{\langle\psi_{\beta}|\psi_{\beta}\rangle}\langle\psi_{\beta}|e^{iHt_{n-1}}\hat{A}e^{-iH(t_{n-1}-t_{n-2})}\hat{A}e^{-iH(t_{n-2}-t_{n-3})} \nonumber \\
   \cdots e^{-iH(t_{2}-t_{1})}\hat{A}e^{-iHt_{1}}\hat{A}|\psi_{\beta}\rangle .
\end{gather}
The free cumulants and classical cumulants are thus calculated according to Eqs. (3) and (7) in the main text.

The time evolution of the state is
calculated by
iterating in real time using Chebyshev polynomial algorithm \cite{Jin10-Chebyshev,Raedt04}.
With similar methods, $|\psi_\beta\rangle$ can be calculated by iterating in imaginary time.
In this paper, we calculate the free and classical cumulants   up to Hilbert space dimension $D = 3^{18}\approx 400000000$, far beyond the limit of ED.

\begin{figure}[t]
	\centering
	\includegraphics[width=1 \linewidth]{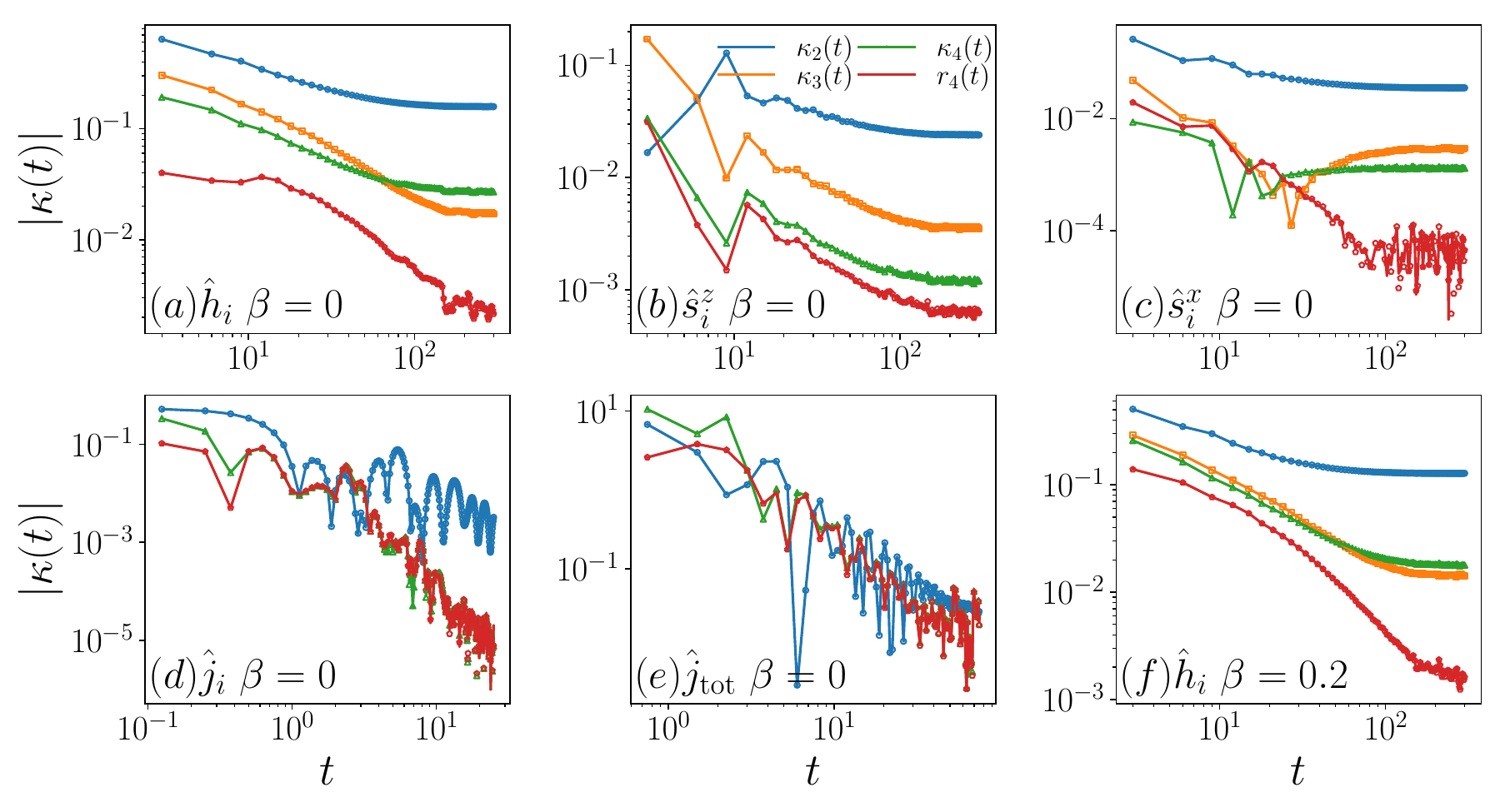}
	\caption{Accuracy check: DQT (open) vs ED (solid line) for free and classical cumulants for $L = 10$ for (a) $\hat{A} = \hat{h}_i, \beta = 0$;(b)$\hat{A} = \hat{s}^z_i, \beta = 0$;
    (c) $\hat{A} = \hat{s}^x_i, \beta = 0$;
    (d) $\hat{A} = \hat{j}_i, \beta = 0$;
    (e) $\hat{A} = \hat{j}_\text{tot}, \beta = 0$;
    (f) $\hat{A} = \hat{h}_i, \beta = 0.2$.
    The DQT results are obtained from an average over $3^8$ different realizations of Haar random states.
    }
	\label{fig:check0}
\end{figure}

\begin{figure}[t]
	\centering
	\includegraphics[width=1 \linewidth]{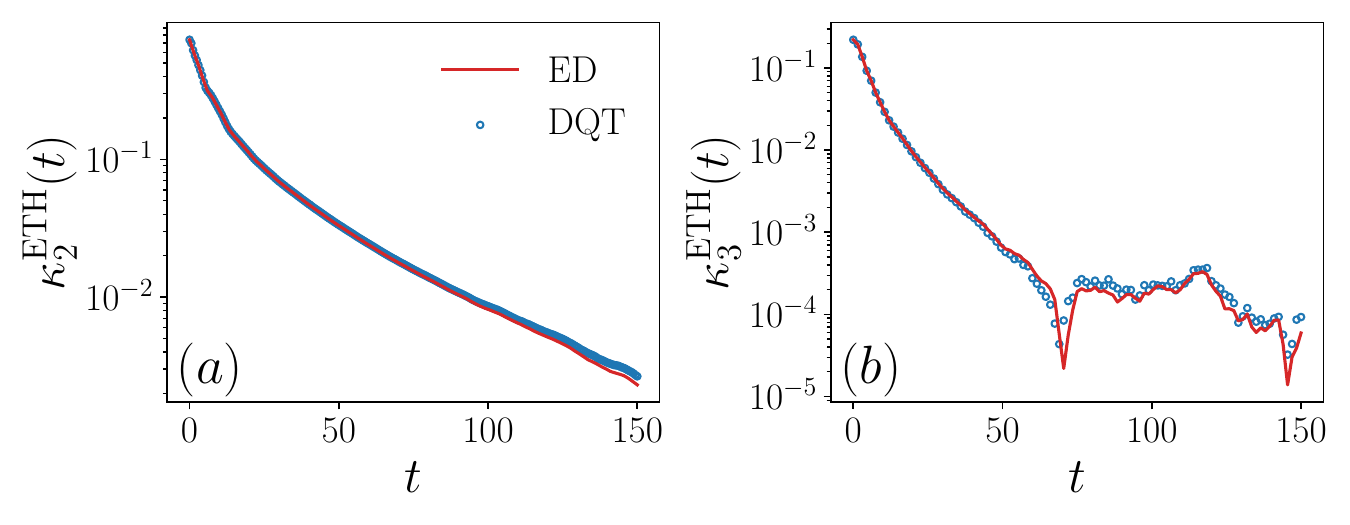}
	\caption{Accuracy check: DQT (open) vs ED (solid line) for ETH free cumulants (a) $\kappa^{\text{ETH}}_2(t)$ and (b) $\kappa^{\text{ETH}}_3(t)$ for $L = 10$ for $\hat{A} = \hat{h}_i, \beta = 0$.   The DQT results are obtained from an average over $3^8$ different realizations of Haar random states.
    }
	\label{fig:check-eth}
\end{figure}
\subsubsection{ETH free cumulants}
The calculation of ETH free cumulants $\kappa^{\text{ETH}}_n$ is significantly more involved than that of free or classical cumulants, due to the fact that they are not fixed by $n$-point correlation functions. It is only feasible through dynamical quantum typicality (DQT) in specific cases, e.g., ${\hat A} = \hat{h}_i$. In the following, we introduce the method to calculate $\kappa^{\text{ETH}}_2$ and $\kappa^{\text{ETH}}_3$ of $\hat{h}_i$.

The second ETH free cumulant reads
\begin{align}
\kappa_{2}^{\text{ETH}}(t)&=\frac{1}{Z}\sum_{i_{1}\neq i_{2}}|A_{i_{1}i_{2}}|^{2}e^{-\beta E_{i_{1}}}e^{-i(E_{i_{1}}-E_{i_{2}})t}\nonumber \\
    & = \langle(\hat{A}(t)-\text{Diag}(\hat{A}))(\hat{A}-\text{Diag}(\hat{A}))\rangle,
\end{align}
where $\text{Diag}(\hat{A})=\sum_{i}A_{ii}|i\rangle\langle i|$. For operator ${\hat A} = \hat{h}_i$, due to the translation invariant symmetry, $\text{Diag}(A) = \frac{\hat H}{L}$.
As a result,
\begin{align}
\kappa_{2}^{\text{ETH}}(t_1, 0)&=\langle (\hat{A}(t_1)-\frac{\hat{H}}{L})(\hat{A}-\frac{\hat{H}}{L}) \rangle \nonumber \\
& =\langle\hat{A}(t_1)\hat{A}\rangle-\frac{2}{L}\langle\hat{A}\hat{H}\rangle+\frac{1}{L^{2}}\langle\hat{H}^{2}\rangle. \label{eq-keth2}
\end{align}

Similarly, the third ETH free cumulant of $\hat{h}_i$ can be written as
\begin{align}
    \kappa_{3}^{\text{ETH}}(t_2,t_1,0) & =\frac{1}{Z}\sum_{i_{1}\neq i_{2}\neq i_{3}}A_{i_{1}i_{2}}A_{i_{2}i_{3}}A_{i_{3}i_{1}}e^{-\beta E_{i_1}} \nonumber \\
    &\cdot e^{-i(E_{i_{1}}-E_{i_{2}})t_{2}-i(E_{i_{2}}-E_{i_{3}})t_{1}} \nonumber \\
    &=\langle(\hat{A}(t_{2})-\frac{\hat{H}}{L})(\hat{A}(t_{1})-\frac{\hat{H}}{L})(\hat{A}-\frac{\hat{H}}{L})\rangle \nonumber \\
    &  =\langle\hat{A}(t_{2})\hat{A}(t_{1})\hat{A}\rangle-\delta_{3}(t_2,t_1,0), \label{eq-keth3}
\end{align}
where
\begin{equation}\label{eq-deth3}
\delta_{3}(t_2,t_1,0)=\frac{\langle\hat{A}(t_{2})\hat{A}(t_{1})\hat{H}\rangle+\langle\hat{A}(t_{1})\hat{A}\hat{H}\rangle+\langle\hat{A}\hat{A}(t_{2})\hat{H}\rangle}{L}-\frac{2\langle\hat{H}^{3}\rangle}{L^{3}}.
\end{equation}
Using Eqs.~\eqref{eq-keth2}, ~\eqref{eq-keth3} and \eqref{eq-deth3}, one can calculate $\kappa_{2}^{\text{ETH}}(t_1,0)$ and $\kappa_{3}^{\text{ETH}}(t_2,t_1,0)$ of $\hat{A} = \hat{h}_i$ via DQT.

\subsubsection{Accuracy check}
To check the accuracy of our DQT results, we compare them with those obtained from exact diagonalization (ED) for a system size of $L = 10$. The comparison is shown in Fig.~\ref{fig:check0} for free and classical cumulants, and in Fig.~\ref{fig:check-eth} for ETH free cumulants.
For all operators considered, the DQT and ED results show nearly perfect agreement.

\end{document}